\documentstyle[12pt]{article}
\newfont{\bg}{cmr10 scaled\magstep4}

\newcommand{\bigzerou}{%
   \smash{\lower1.7ex\hbox{\bg 0}}}
\begin{document}
\title{
\begin{flushright}
  \begin{minipage}[b]{5em}
    \normalsize
      UT-HEP-734-95\\
  \end{minipage}
\end{flushright}
On Quantum Cohomology Rings \\
for Hypersurfaces in $CP^{N-1}$}
\author{Masao Jinzenji\\
\\
\it Department of Physics, University of Tokyo\\
\it  Bunkyo-ku, Tokyo 113, Japan}
\maketitle
\begin{abstract}
Using the torus action method, we construct 
one variable polynomial representation of 
quantum cohomology ring for  degree $k$ hypersurface in 
$CP^{N-1}$ . The results interpolate the well-known result of 
$CP^{N-2}$ model and the one of Calabi-Yau hypersuface in 
$CP^{N-1}$. We find in $k\leq N-2$ case, principal 
relation of this ring have very simple form 
compatible with toric compactification of moduli space 
of holomorphic maps from $CP^{1}$ to $CP^{N-1}$.
\footnote{ e-mail 
address: jin@danjuro.phys.s.u-tokyo.ac.jp}  
\end{abstract}

\section{Introduction}
 In recent years, study of topological sigma model has made a
great progress. In my point of view, these models are classified 
into following types.\\

A. topological sigma model (pure matter theory) with the \\
\quad 1. target space  $M$ with $c_{1}(T'M) > 0$\\
\quad 2. target space  $M$ with $c_{1}(T'M)=0$ (Calabi-Yau manifold)\\
\quad 3. target space  $M$ with $c_{1}(T'M) < 0$\\

B. topological sigma model coupled to topological gravity with the \\
\quad 1. target space  $M$ with $c_{1}(T'M) > 0$\\
\quad 2. target space  $M$ with $c_{1}(T'M)=0$ (Calabi-Yau manifold)\\
\quad 3. target space  $M$ with $c_{1}(T'M) < 0$\\

$T'M$ denotes holomorphic part of tangent bundle of K\"ahler 
manifold $M$.
  
For cases A.2 and B.2, we can solve these models using mirror 
symmetry \cite{candelas2},\cite{ooguri} and \cite{mnmj}. We may 
also consider the case when the target manifold $M$ is Fano variety, 
i.e., $c_{1}(T'M)\geq 0$.
In \cite{km} and \cite{dub}, it was  shown that models coupled to 
gravity (small phase space, tree level) can be solved imposing 
associativity condition of operator algebra. In fact, calculation 
of amplitudes for general Fano variety by using this method is 
tedious, but we can't deny the effectiveness of this approach. 
On the other hand, when target space is $CP^{1}$, the simplest 
Fano variety, the matrix integral representation of partition 
function of the model coupled to gravity (large phase space,
all genus) was given in \cite{ey} by constructing Lax operator 
formalism of flows induced from insertion of all BRST closed operators.
Then we can naturally ask whether we can construct Lax operator 
representation of  sigma models on general Fano varieties. But 
we don't pursue this problem in this paper. 

In this paper we consider the case  A.1, topological sigma models 
on Fano varaeties without coupling to gravity. 
In this case, Vafa and Intriligator \cite{fusion} conjectured 
quantum cohomology rings that correspond to the solutions of sigma 
models on Grassmanianns. They first considered Landau-Ginzburg 
potential $W(X)$ that determines the relations (ideal) of classical 
cohomology ring of Grassmannians. For example, $W(X)$ in  $CP^{N}$ model 
 is given by 
\begin{equation}
W^{CP^{N}}(X)= \frac{X^{N+2}}{N+2}
\label{pot}
\end{equation}
where $X$ represents the K\"ahler form of $CP^{N}$, the generator 
of $H^{*}(CP^{N},C)$.
Then  the relation of $H^{*}(CP^{N},C)$ is obtained from 
$\partial_{X}W(X)=0$
\begin{equation}
\partial_{X}W^{CP^{N}}(X)=X^{N+1}=0.
\label{lgrel}
\end{equation}
From this we can determine ring structure of  $H^{*}(CP^{N},C)$.
Then they argued that relations of the 
quantum cohomology ring are generated 
by Landau-Ginsburg potential perturbed along the direction 
of K\"ahler form, 
\begin{equation}
W^{CP^{N}}_{q}(X)= \frac{X^{N+2}}{N+2}-Xq
\label{qpot}
\end{equation} 
where $q (:= e^{-t})$ is the  deformation parameter which counts 
the degree of instanton (holomorphic maps from Riemann surface 
to $CP^{N}$). Relation of quantum cohomology ring is derived in 
the same way as the classical case,
\begin{equation}
\partial_{X}W_{q}^{CP^{N-1}}(X)=X^{N+1}-q=0.
\label{qrel}
\end{equation}    
For a general Grassmannian $Gr(N,N+M)$ whose cohomology ring has 
$N$ generators, the situation is the same. Let $W_{N+M+1}^{(N)}(X_{1},
\cdots,X_{N})$ \footnote{Let $X^{(N)}(t) = \sum_{i=0}^{N}X_{i}t^{i}$.
Then $W_{N+M+1}^{(N)}(X_{1},\cdots,X_{N})$ is given by $t^{N+M+1}$ 
coefficient of $W^{(N)}(t)= -\log(X^{(N)}(-t))=\sum_{i}W_{i}^{(N)}(X_{1},
\cdots,X_{N}) t^{i}$} be the Landau-Ginzburg potential for  $Gr(N,N+M)$, 
then  Landau-Ginzburg potential of quantum cohomology ring is 
\begin{equation}
W_{q}^{Gr(N,N+M)}(X_{1},\cdots,X_{N}):= W_{N+M+1}^{(N)}(X_{1},\cdots,X_{N})+(-1)^NqX_{1}
\label{qgr}
\end{equation}
and relations of this ring are given by $\partial_{X_{i}}W_{q}^{Gr(N,N+m)}=0$.
 One can evaluate correlation function of these models from 
the residue formula,
\begin{equation}
\langle \prod_{j=1}^{m}{\cal O}_{A_{j}}(z_{j})\rangle
=\oint_{C_{\infty}}\cdots \oint_{C_{\infty}} 
\frac{\prod_{j=1}^{m}f_{A_{j}}(X)}
{\prod_{i=1}^{N}(\partial_{X_{i}}W_{q}^{Gr(N,N+M)}(X))}
dX_{1}\cdots dX_{N}
\label{res}
\end{equation}
where $f_{A_{i}}(X)$ represents $A_{i}\in H^{*}(Gr(N,N+M),C)$ in 
Landau-Ginzburg representation of $H^{*}(Gr(N,N+M),C)$ 
and ${\cal O}_{A_{i}}$ denotes BRST-closed 
operator of sigma model induced from $A_{i}$. In this case, $f_{A_{i}}(X)$
receives no quantum correction. Correlation functions  evaluated 
from (\ref{res}) take integer values and are non-vanishing only if the  
topological selection rule of sigma model is satisfied.   

In this construction, geometrical aspect of sigma models, i.e., counting 
of instantons from Riemann surface to target space $M$ is not clear. 
But let us  assume the fusion rule of correlation function 
of pure matter theory which can be derived from taking the position of 
operator insertion points $z_{1}, z_{2}$ in an infinitesimally
small disc, or equivalently from putting  $z_{1}, z_{2}$ on one 
component of a stable curve with one branch point and $z_{3}, \cdots, z_{k}$
on the other. 
\begin{equation}
\langle \prod_{j=1}^{m}{\cal O}_{A_{j}}(z_{j})\rangle=
\langle{\cal O}_{A_{1}}(z_{1}){\cal O}_{A_{2}}(z_{2})
    {\cal O}_{A_{a}}(z_{s1})\rangle\eta^{ab}
\langle{\cal O}_{A_{b}}(z_{s2})\prod_{j=3}^{m}{\cal O}_{A_{j}}(z_{j})
    \rangle
\label{ifus}
\end{equation}
In (\ref{ifus}), $z_{s1}$ and $z_{s2}$ are  
the positions of the double singularities on 
each component and $\eta^{ab}$ is the flat metric defined from 
two point functions of the model which receives no quantum correction.
\begin{eqnarray}
\eta_{ab}:&=& \langle{\cal O}_{A_{a}}(z_{1}){\cal O}_{A_{b}}(z_{2})\rangle
          = \int_{M}A_{a}\wedge A_{b}\nonumber\\
 \eta^{ab}\eta_{bc}&=&\delta^{a}_{c}
\label{prop}
\end{eqnarray}
This relation naturally leads us to consider the algebra defined 
by
\begin{equation}
{\cal O}_{A_{1}}\cdot{\cal O}_{A_{2}}=\langle{\cal O}_{A_{1}}{\cal O}_{A_{2}}
    {\cal O}_{A_{a}}\rangle\eta^{ab}{\cal O}_{A_{b}}. 
\label{adef}
\end{equation}
Note that we can regard this definition as the generalized form 
of multiplication rule of classical cohomology algebra 
\begin{equation}
A_{1}\cdot A_{2}= (\int_{M}A_{1}\wedge A_{2}\wedge A_{a})\eta^{ab}A_{b}.
\label{cdef}
\end{equation}
Then the geometrical evaluation of three point functions should 
reproduce the quantum cohomology algebra of Grassmannians and coincide 
with the result of Landau-Ginzburg approach. This line of discussion was 
done by Bertram and  Ruan et al \cite{bert},\cite{ruan}. 

In this paper we treat sigma models on degree $k$ hypersurface in $CP^{N-1}$
, $M_{N}^{k}$. $M_{N}^{k}$ has a unique K\"ahler form $e$ that descends from 
the K\"ahler form of $CP^{N-1}$. We can consider $e$ as the generator of the  
subring of $H^{*}(M_{N}^{k},C)$, which consists of $e^{k} (k=1,2,\cdots,
N-2)$.
 We denote it as  $H^{*}_{e}(M_{N}^{k},C)$.
Classically,  $H^{*}_{e}(M_{N}^{k},C)$ has the following ring structure.
\begin{equation}
e^{N-1}=0
\label{erel}
\end{equation} 
One can see that from (\ref{lgrel}), (\ref{erel}) is the same as 
the relation of $H^{*}(CP^{N-2},C)$. Of course, we can also derive 
the same ring structure by using classical three point functions 
and metric,
\begin{equation}
e^{\alpha}\cdot e^{\beta}=(\int_{M_{N}^{k}}e^{\alpha}\wedge e^{\beta}
\wedge e^{\gamma})\eta^{\gamma \delta}e^{\delta}
\label{emul}
\end{equation}
where
\begin{eqnarray}
\int_{M_{N}^{k}}e^{\alpha}\wedge e^{\beta}
\wedge e^{\gamma}&=&k\cdot\delta_{\alpha+\beta+\gamma,N-2}\nonumber\\
\eta^{\alpha\beta}&=&\frac{1}{k}\delta_{\alpha+\beta,N-2}.
\label{strpro}
\end{eqnarray}  
Then our assertion is that quantum version 
of  $H^{*}_{e}(M_{N}^{k},C)$ (we denote it as $H^{*}_{q,e}(M_{N}^{k},C)$) 
is given by 
\begin{equation}
{\cal O}_{e^{\alpha}}\cdot{\cal O}_{e^{\beta}}=
\langle {\cal O}_{e^{\alpha}}{\cal O}_{e^{\beta}}{\cal O}_{e^{\gamma}}
\rangle\eta^{\gamma\delta}{\cal O}_{e^{\delta}}.
\label{qerel}
\end{equation}
Degree counting parameter $q$ reveals itself in 
$\langle {\cal O}_{e^{\alpha}}{\cal O}_{e^{\beta}}{\cal O}_{e^{\gamma}}
\rangle$.
This is equivalent to the assumption of restricted fusion rule,
\begin{equation}
\langle \prod_{j=1}^{m}{\cal O}_{e^{\alpha_{j}}}(z_{j})\rangle=
\langle{\cal O}_{e^{\alpha_{1}}}(z_{1}){\cal O}_{e^{\alpha_{2}}}(z_{2})
    {\cal O}_{e^{a}}(z_{s1})\rangle\eta^{ab}
\langle{\cal O}_{e^{b}}(z_{s2})\prod_{j=3}^{m}
{\cal O}_{e^{\alpha_{j}}}(z_{j})
    \rangle.
\label{rfus}
\end{equation}
In (\ref{rfus}), we assumed that n-point correlation function with 
$(n-1)$ insertions of ${\cal O}_{e^{\alpha}}$'s and 1 insertion 
from ${\cal O}_{A} (A\notin H^{*}_{e}(M_{N}^{k},C))$ vanishes.
If $N$ is odd, (i.e., $\mbox{dim}(M_{N}^{k})$ is odd) justification 
of this assumption is easy. Because there is no other analytic class
 in $ H^{*}(M_{N}^{k},C)$ (with equal holomorphic and anti-holomorphic 
form degree) except for elements of $ H^{*}_{e}(M_{N}^{k},C)$. 
Then the above correlation function vanishes from topological selection 
rule.
In case N is even, analytic classes which are not included in 
$ H^{*}_{e}(M_{N}^{k},C)$ do appear in  $H^{\frac{N}{2}-1,\frac{N}{2}-1}
(M_{N}^{k},C)$, but numerical results of 
this paper  makes no contradiction with (\ref{rfus}).

Then what remains to consider is the evaluation of three point functions.
These are computed  from two facts :\\ 
1. Three point functions of pure 
matter theory  and  of theory coupled to gravity coincide. \\
2. Three point 
functions of theory coupled to gravity are evaluated by using the torus 
action method \cite{tor}.\\
 In this way, we determined the ring structure of 
$ H^{*}_{q,e}(M_{N}^{k},C)$ and find the quantum generalization of 
the relation (\ref{erel}) if $k$ is no more than $N-2$.
\begin{equation}
{\cal O}_{e}^{N-1}-k^{k}{\cal O}_{e}^{k-1}q=0
\label{new}
\end{equation}
By setting ${\cal O}_{e}:= X$, we can regard (\ref{new}) as generalization
of (\ref{qrel}), 
\begin{equation}
\partial_{X}W_{q}^{M_{N}^{k}}=X^{N-1}-k^{k}X^{k-1}q:=f_{rel}(X)
\label{rewrite}
\end{equation} 
which reduces to the result of $CP^{N-2}$ model in $k=1$ case.
(\ref{new}) tells us that for $k\geq 2$ case, the direction of 
deformation is no longer the one of K\"ahler operator, and 
${\cal O}_{e^{\alpha}} (\alpha\geq 2)$ indeed receives quantum correction 
in this case.
\begin{eqnarray}
{\cal O}_{e^{\alpha}}&=& X^{\alpha}-\sum_{d=1}^{[\frac{\alpha}{N-k}]}
\gamma_{\alpha,d}^{N,k}X^{\alpha-(N-k)d}q^{d}:=f_{\alpha}(X)
\nonumber\\
&&(\alpha=2,\cdots, N-2)
\label{corre}
\end{eqnarray}
 We found some curious relations among $\gamma_{\alpha,d}^{N,k}$'s 
which seems to suggest that higher quantum corrections ($d\geq 2$)
are written in terms of $\gamma_{\alpha,1}^{N,k}$'s, i.e., degree 
1 instanton corrections. This is natural because relation (\ref{rewrite})
receives corrections only from degree 1 sector. 

With these results we obtain the residue formula of correlation 
functions with full insertion of elements of $H^{*}_{q,e}(M_{N}^{k},C)$.
\begin{equation}
\langle\prod_{j=1}^{m}{\cal O}_{e^{\alpha^{j}}}(z_{j})\rangle
=k\cdot\oint_{\infty}\frac{\prod_{j=1}^{m}f_{\alpha_{j}}(X)}{f_{rel}(X)}dX
\label{gres}
\end{equation}
This is the generalization of (\ref{res}).

The relation (\ref{new}) has some geometrical meaning. Using (\ref{gres}),
we can easily see 
\begin{equation}
\langle\prod_{j=1}^{(N-2)+(N-k)d}{\cal O}_{e}(z_{j})\rangle
=k^{kd+1}q^{d}.
\label{motiva}
\end{equation}
In \cite{mnmj}, we found that correlation function 
$\langle\prod_{j=1}^{N-2}{\cal O}_{e}(z_{j})\rangle =
\sum_{d=0}^{\infty}\langle\prod_{j=1}^{N-2}
{\cal O}_{e}(z_{j})\rangle_{d}q^{d}$ of sigma model on Calabi-Yau hyper 
surface in $CP^{N-1}\;\; (M_{N}^{N})$ has the following structure 
\begin{equation}
\langle \prod_{i=1}^{N-2}{\cal O}_{e}(z_{i})\rangle_{d} = (N^{dN+1}- 
          (\mbox{correction terms})),
\label{cy} 
\end{equation}
using mirror symmetry. We asserted there that the top term $(N^{dN+1})$ can be 
evaluated from simple compactification of moduli space 
${\cal M}_{0,d}^{CP^{N-1}}$ of instantons (holomorphic maps) from 
$CP^{1}$ to $CP^{N-1}$ (We compactified ${\cal M}_{0,d}^{CP^{N-1}}$
into $CP^{(d+1)N-1}$) and that correction terms appear as the effect of 
boundary components added in the process of compactification. 
This argument is shown to correspond to the result of gauged linear 
sigma model in \cite{mp}. (\ref{motiva}) tells us that there are no
correction terms in $k\leq N-2$ case. This fact can be explained 
from the above point of view. Boundary components are irrelevant 
in this case in view of the  dimensional counting.


In section 2, we review fundamental facts of topological sigma 
model (A-model) for pure matter case, introduce the notion 
of quantum cohomology algebra and discuss its applications 
to pure A-model.
   
 In section 3, we briefly explain the strategy of the construction of 
quantum cohomology ring of $M_{N}^{k}$. In section 4, we reformulate 
this ring as the polynomial algebra of ${\cal O}_{e}$ and discuss 
its characteristic structure when $k\leq N-2$. In section 5, we 
try to understand geometrical reason of the bound $k\leq N-2$ and 
explain why under this bound boundaries of $CP^{d(N+1)-1}-
{\cal M}_{0,d}^{CP^{N-1}}$ are irrelevant in integration of forms.

All numerical results used in this paper are collected at the end 
of this paper.

\section{Topological Sigma Model (A-model)}
Topological sigma model is constructed from twisting 
$N=2$ super symmetric sigma model that describes maps from 
Riemann surface $\Sigma$ to K\"ahler manifold $M$
(In this paper we limit genus of Riemann surface to 0). 
Then we have the following Lagrangian.
\begin{equation}
L= 2t\int_{\Sigma}d^{2}z( \frac{1}{2}g_{i\bar{j}}(\partial_{z}\phi^{i}
\partial_{\bar{z}}\phi^{\bar{j}}+
\partial_{z}\phi^{\bar{j}}
\partial_{\bar{z}}\phi^{i})
+i\psi_{z}^{\bar{i}}D_{\bar{z}}\chi^{i}g_{i\bar{i}}
+i\psi_{\bar{z}}^{i}D_{z}\chi^{\bar{i}}g_{i\bar{i}}
-R_{i\bar{i}j\bar{j}}\psi^{i}_{\bar{z}}\psi^{\bar{i}}_{z}
\chi^{j}\chi^{\bar{j}})
\label{lag}
\end{equation}
where $\phi$ is the map from $CP^{1}$ to $M$, the
only bosonic degrees of freedom of this model, and $\chi$ is 
fermionic ghost fields with ghost number $1$ and $\psi$ 
are anti-ghosts with ghost number $-1$. 
This Lagrangian is invariant under the BRST-transformation,
\begin{eqnarray}
    \delta\phi^{i}&=&i\alpha\chi^{i},\;
    \delta\phi^{\bar{i}}=i\alpha\chi^{\bar{i}}\nonumber\\
     \delta\chi_{i}&=&0,\;
     \delta\chi_{\bar{i}}=0\nonumber\\
     \delta\psi^{\bar{i}}_{z}&=&-\alpha\partial_{z}\phi^{\bar{i}}
            -i\alpha\chi^{\bar{j}}\Gamma^{\bar{i}}_{\bar{j}\bar{m}}
             \psi^{\bar{m}}_{z}\nonumber\\
     \delta\psi^{i}_{\bar{z}}&=&-\alpha\partial_{\bar{z}}\phi^{i}
            -i\alpha\chi^{j}\Gamma^{i}_{jm})
             \psi^{m}_{\bar{z}}.
\label{BRST}
\end{eqnarray}
We define the generator of this transformation as $Q$ ,i.e., $Q$ acts 
as $\delta V=-\alpha\{Q,V\}$ for any field $V$. This transformation 
is nilpotent and we only have to consider BRST-closed operator as 
observables of this theory.

BRST-closed observables are constructed from closed form on $M$.
Let $A=A_{i_{1}i_{2}\cdots i_{k}}d\phi^{i_{1}}\wedge\cdots\wedge
d\phi^{i_{k}}$ be an $k$-form on $M$ and we define a corresponding 
operator ${\cal O}_{A}:=A_{i_{1}i_{2}\cdots i_{k}}\chi^{i_{1}}\cdots
\chi^{i_{k}}$. Then from BRST-transformation rules, we have  
\begin{equation}
\{Q,{\cal O}_{A}\}= -{\cal O}_{dA}.
\label{obs}
\end{equation}
This tells us that we can construct BRST-closed observable ${\cal O}_{A}$
from the element $A$ of de Rham cohomology ring $H^{*}(M)$.

Of course, Lagrangian $L$ satisfies $\{Q,L\}=0$. Moreover we can rewrite
$L$ modulo $\psi$ equation of motion in the following form.
\begin{eqnarray}
L& =& it\int_{\Sigma}d^{z}\{Q,V\}+t\int_{\Sigma}\Phi^{*}(e)
\label{div}
\end{eqnarray}
where
\begin{eqnarray}
V&=&g_{i\bar{j}}(\psi_{z}^{\bar{i}}\partial_{\bar{z}}\phi^{j}+
                 \partial_{z}\phi^{\bar{i}}\psi^{j}_{\bar{z}})\\
\int_{\Sigma}\Phi^{*}(e)&=&\int_{\Sigma}(\partial_{z}\phi^{i}
                           \partial_{\bar{z}}\phi^{\bar{j}}g_{i\bar{j}}
                           -\partial_{\bar{z}}\phi^{i}
                           \partial_{z}\phi^{\bar{j}}g_{i\bar{j}})
\label{rel}
\end{eqnarray}
$\int_{\Sigma}\Phi^{*}(e)$ is the integral of pull-back of the K\"ahler
form $e$ of $M$ and equals  $\Phi_{*}(\Sigma)\cap PD(e)$ where $PD(W)$ 
denotes Poincare dual of $W\in H^{*}(M)$. We call 
 $\Phi_{*}(\Sigma)\cap PD(e)$ as the degree of maps and divide the phase
space into sectors $B_{d}$ with fixed degree $d$. In $B_{d}$, the second
term of (\ref{div}) is fixed to $td$. 
Then since the first term of (\ref{div}) is BRST-exact, path-integral 
is invariant under the variation of the coupling constant. We can easily 
check this by taking infinitesimal variation of $t$. And we can take
weak coupling limit $t\rightarrow \infty$. In this limit, the saddle point
approximation of path integral becomes exact and the saddle point is given 
by the fixed point of BRST-transformation,
\begin{eqnarray}
\chi^{i}&=0,\quad\chi^{\bar{i}}&=0\nonumber\\
\psi_{z}^{\bar{i}}&=0,\quad \psi_{\bar{z}}^{i}&=0\nonumber\\
\partial_{z}\phi^{\bar{i}}&=0,\quad\partial_{\bar{z}}\phi^{i}&=0 
\label{fixed}
\end{eqnarray}
Especially, the saddle point of bosonic degrees of freedom is given by
holomorphic map from $CP^{1}$ to $M$. This saddle point has moduli and 
we denote this moduli space as ${\cal M}_{0,d}^{M}$ where is $d$  is the 
degree of  holomorphic maps.
We can count the dimension of ${\cal M}_{0,d}^{M}$ by Riemann-Roch 
theorem,
\begin{eqnarray}
\mbox{dim}({\cal M}_{0,d}^{M})&=&
\mbox{dim}(H^{0}(\Sigma,f^{*}(T'M)))\nonumber\\
     &=& \mbox{dim}(M)+ d\cdot c_{1}(T'M)                           
         +\mbox{dim}(H^{1}(\Sigma,f^{*}(T'M)))\\
  a_{d}&:=&\mbox{dim}(M)+ d\cdot c_{1}(T'M)
\label{dim}
\end{eqnarray}
where $f$ denotes holomorphic map from $\Sigma$ to $M$ 
considered  as the point of ${\cal M}_{0,d}^{M}$.  
 Correspondingly, $\chi$ zero modes $(\chi_{0})$, the number of which is 
equal to $\mbox{dim}(H^{0}(\Sigma,f^{*}(T'M)))$, and $\psi$
zero modes ($\psi_{0}$), the number of which is equal to 
$\mbox{dim}
(H^{1}(\Sigma,f^{*}(T'M)))$ occur.
Then by integrating over non-zero modes, we obtain effective 
Lagrangian $L_{eff}$ containing $\psi_{0}$ and $\chi_{0}$.

Since $L_{eff}$ conserves ghost number,
we have to compensate 
this ghost number anomaly by external operator insertions.
To make situations  simpler, we treat observables induced from 
analytic cohomology elements $A \in H^{i,i}(M,C)$. We define 
ghost number of ${\cal O}_{A}$ as $\mbox{dim}_{C}(A)=i$. 
Then cancelation condition of ghost number anomaly reduces to 
\begin{eqnarray}
\langle \prod_{j=1}^{k}{\cal O}_{A_{j}}(z_{j})\rangle_{d}
&=&\int d{\cal M}_{0,d}^{M} d\chi_{0} d\psi_{0} 
\exp(- L_{eff})
\prod_{j=1}^{k}{\cal O}_{w_{j}}(z_{j})\neq 0\nonumber\\
&\Longrightarrow&
\sum_{j=1}^{k}\mbox{dim}_{C}(A_{k})=
\mbox{dim}(H^{0}(\Sigma,f^{*}(T'M)))
-\mbox{dim}(H^{1}(\Sigma,f^{*}(T'M)))\nonumber\\
&=& \mbox{dim}(M)+ d\cdot c_{1}(T'M),
\label{sel}
\end{eqnarray} 
This is the topological selection rule.

And with this selection rule, we can rewrite 
$\langle \prod_{j=1}^{k}{\cal O}_{A_{j}}(z_{j})\rangle$ into the following 
form.
\begin{eqnarray}
\langle \prod_{j=1}^{k}{\cal O}_{A_{j}}(z_{j})\rangle&=&
\sum_{d=0}^{\infty}\delta_{(\sum_{j=1}^{k} dim_{C}(A_{j}),a_{d})}
\int d{\cal M}_{0,d}^{M} d\chi_{0} d\psi_{0} 
\exp(- L_{eff})
\prod_{j=1}^{k}{\cal O}_{A_{j}}(z_{j})\cdot q^{d}\nonumber\\
&:=&\sum_{d=0}^{\infty}\delta_{(\sum_{j=1}^{k} dim_{C}(A_{j}),a_{d})}
\langle \prod_{j=1}^{k}{\cal O}_{A_{j}}(z_{j})\rangle_{d}\cdot q^{d}
\nonumber\\
q&:=& e^{-t}
\label{dec}
\end{eqnarray}
With this set up, we evaluate path-integral.  Note that remaining
degrees of freedom is moduli space of holomorphic maps from 
$\Sigma$ to $M$ and $\psi$ and $\chi$ zero modes.
First, we consider the  
generic case when \\
$\mbox{dim}(H^{1}(\Sigma,f^{*}(T^{1,0}M)))=0$.
In this case, $L_{eff.}$ equals zero because it conserves 
ghost number. 
 By taking $A_{j}\in H^{i,i}(M,C)$ as the form which 
has delta function
 support on $PD(A_{j})$, we pick up holomorphic maps 
which satisfies the following condition in integration of
moduli space.
\begin{equation}
f(z_{j})\in PD(A_{j}) \quad (j=1,\cdots, k)
\label{geoc}
\end{equation}
These conditions impose $\mbox{dim}_{C}(A_{j})$ dimensional 
constraint on ${\cal M}_{0,d}^{M}$ for each $j$ because we use 
$(\mbox{dim}_{C}(A_{j})-1)$ degrees of freedom to make $f(CP^{1})$ 
intersect $PD(A_{j})$ and $1$ degrees of freedom to make 
$f(z_{j})\in PD(A_{j})$. These constraints kill all the moduli degrees 
of freedom as we can easily see from (\ref{sel}). Remaining  fermionic 
fields are balanced by the measure of zero modes in the Grassmann 
integrand, and correlation function results in
\begin{equation}
\langle \prod_{j=1}^{k}{\cal O}_{A_{j}}(z_{j})\rangle_{d}=
{}^{\sharp}\{f:\Sigma\stackrel{hol.}{\mapsto}M|\; \mbox{deg}(f)=d,\;\;
f(z_{j})\in PD(A_{j}) \quad (j=1,\cdots, k)\}
\label{cor1}
\end{equation} 
We can rewrite (\ref{cor1}) into more compact form using the evaluation
map
\begin{eqnarray}
\phi_{j}: {\cal M}_{0,d}^{M}&\mapsto& M\nonumber\\
               f\in  {\cal M}_{0,d}^{M}&\mapsto& f(z_{j}),
\label{ev}
\end{eqnarray}
as follows.
\begin{equation}
\langle \prod_{j=1}^{k}{\cal O}_{A_{j}}(z_{j})\rangle_{d}=
\int_{ {\cal M}_{0,d}^{M}}\bigwedge_{j=1}^{k}\phi_{j}^{*}(A_{j})
\label{cor2}
\end{equation}
 
 Next, we consider the non-generic case when 
$\mbox{dim}_{C}(H^{1}(CP^{1},f^{*}(T'M)))\neq 0$. From Kodaira-Serre 
duality, we have 
\begin{equation}
(H^{1}(CP^{1},f^{*}(T'M)))^{*}= H^{0}(CP^{1},K_{CP^{1}}\otimes f^{*}(T'M))
           = H^{0}(CP^{1},{\cal O}_{CP^{1}}(-2)\otimes f^{*}(T'M)).
\label{kod}
\end{equation}
On the other hand, since  
$c_{1}(f^{*}(T'M))=f^{*}(c_{1}(T'M))=d\cdot c_{1}(T'M)$ 
and $f^{*}(T'M)$ is rank $\mbox{dim}_{C}(M)$ bundle on $CP^{1}$, $f^{*}(T'M)$
is decomposed into a direct sum of line bundles on $CP^{1}$ as 
follows.
\begin{eqnarray}
f^{*}(T'M)&=&\bigoplus_{j=1}^{{dim}_{C}(M)}{\cal O}_{CP^1}(a_{j})\;\;\quad\quad
\quad(\sum_{j=1}^{dim_{C}(M)}a_{j}=d\cdot c_{1}(T'M)\quad)\nonumber\\
&&\mbox{dim}(H^{0}(CP^{1},f^{*}(T'M)))=\sum_{a_{j}\geq 0}(a_{j}+1)
\label{bundle}
\end{eqnarray}
From (\ref{kod}) and (\ref{bundle}), we obtain 
\begin{equation}
{\cal O}(-2)\otimes f^{*}(T'M)=
\bigoplus_{j=1}^{{dim}_{C}(M)}{\cal O}_{CP^{1}}(-2-a_{j})
\label{h1}
\end{equation}
and
\begin{equation}
\mbox{dim}(H^{1}(CP^{1},f^{*}(T'M)))=\sum_{-2-a_{j}\geq 0}(-1-a_{j})
\label{dimh1}
\end{equation}
Then we assume the following.

{\bf Assumption}
{ \it For an irreducible map $f$ (having no multiple cover map component), 
$\mbox{dim}(H^{1}(CP^{1},f^{*}(T'M)))$ equals 0.  
And $H^{0}(CP^{1},f^{*}(T'M))$ has at least one ${\cal O}_{CP^{1}}(2)$
component that  corresponds to an automorphism  of $f(CP^{1})$ and minimal 
${\cal O}_{CP^{1}}(-m)$ ($m$ positive integer) insertion.} 

Under this assumption, we obtain two cases when $f^{*}(T'(M))$ has 
${\cal O}_{CP^{1}}(-m)$ ($m$ positive integer) insertion.
\begin{eqnarray}
\mbox{case}
\; 1\quad f^{*}(T'M)&=&{\cal O}_{CP^{1}}(2)\oplus{\cal O}_{CP^{1}}(-1)
\oplus{\cal O}_{CP^{1}}(-1)\oplus
{\cal O}_{CP^{1}}^{dim_{M}-3} \nonumber\\
&&(c_{1}(T'M)=0,\;d:\mbox{arbitrary})\nonumber\\
\mbox{case}
\; 2\quad f^{*}(T'M)&=&{\cal O}_{CP^{1}}(2)\oplus{\cal O}_{CP^{1}}(-1)
\oplus
{\cal O}_{CP^{1}}^{dim_{M}-2} \nonumber\\ 
&&(c_{1}(T'M)=1,\;d=1)
\label{case}
\end{eqnarray}
The reason why we pick up the above two cases is that reduced map 
$\tilde{f}\;(=f\circ\varphi$ ,$\varphi$ is degree $n$ holomorphic map 
from $CP^{1}$ to $CP^{1}$) indeed has non zero $H^{1}$. 
For these cases, $\tilde{f}^{*}(T'M)$'s are 
\begin{eqnarray}
\mbox{case}
\; 1\quad \tilde{f}^{*}(T'M)&=&
{\cal O}_{CP^{1}}(2n)\oplus{\cal O}_{CP^{1}}(-n)
\oplus{\cal O}_{CP^{1}}(-n)\oplus
{\cal O}_{CP^{1}}^{dim_{M}-3} \nonumber\\
\mbox{case}\; 2\quad \tilde{f}^{*}(T'M)
&=&{\cal O}_{CP^{1}}(2n)\oplus{\cal O}_{CP^{1}}(-n)
\oplus
{\cal O}_{CP^{1}}^{dim_{M}-2},  
\label{mcase}
\end{eqnarray} 
and $\mbox{dim}H^{1}$'s are counted as follows.
\begin{eqnarray} 
\mbox{case}\; 1\quad \mbox{dim}(H^{1}(CP^{1},\tilde{f}^{*}(T'M)))&=&
2n-2\nonumber\\
\mbox{case}\; 2\quad \mbox{dim}(H^{1}(CP^{1},\tilde{f}^{*}(T'M)))
&=& n-1
\label{h1case}
\end{eqnarray}
In such cases, these $H^{1}$'s are stable and we can integrate out 
corresponding $\psi_{0}$ with weight $\exp(-L_{eff})$. In \cite{hori}
and \cite{ap}, it was shown that these integration results in the top 
Chern class of obstruction bundle ${\cal F}$ ($H^{1}$ bundle ) 
on ${\cal M}_{0,d,n}^{M}$
(we denote ${\cal M}_{0,d,n}^{M}$ as  components of moduli 
space of reduced maps of degree $d$ with $n$ multiple cover). 
Then we obtain following formula.
\begin{equation}
\langle\prod_{j=1}^{m}{\cal O}_{A_{j}}(z_{j})\rangle_{d}
=\sum_{n|d}\int_{{\cal M}_{0,d,n}^{M}}c_{T}({\cal F})\wedge
(\bigwedge_{j=1}^{m}\phi_{j}^{*}(A_{j}))
\label{general}
\end{equation}
Of course, the generic case is included in (\ref{general}), if we consider 
${\cal F}$ as rank 0 bundle in this case. In case 1, we can further 
reduce ${\cal F}$ as a bundle on ${\cal M}_{0,n}^{CP^{1}}$ and derive 
well-known multiple cover formula for models on Calabi-Yau manifolds. 
In case 2, we cannot argue heuristically that ${\cal F}$ as a bundle 
on ${\cal M}_{0,n}^{CP^{1}}$ because in this case, 
$\mbox{dim}({\cal M}_{0,n}^{CP^{1}})-\mbox{dim}
((H^{1}(CP^{1},\tilde{f}^{*}(T'M))))= n+2$ and we have nontrivial 
mixture of the top Chern class and evaluation map contribution. 
We also have to note that in this case, non trivial contribution 
should come only from $d=n$ part. Anyway further analysis is needed
\cite{j2}.  
 
Lastly, we explain the notion of the quantum cohomology ring on $M$ as 
the extension of the classical cohomology rings on $M$.
Ring structure of classical cohomology ring on $M$ is determined 
by the classical three point function $C_{ijk}$ and the metric $\eta^{ij}$ 
defined by
\begin{eqnarray}
C_{ijk}&:=& \int_{M}A_{i}\wedge A_{j}\wedge A_{k},
\quad \eta_{ij}:=\int_{M}A_{i}\wedge A_{j}\nonumber\\
\eta_{ij}\eta^{jk}&=&\delta_{i}^{k}.
\label{cc}
\end{eqnarray} 
In this definition multiplication rules are written as follows
\begin{equation}
A_{i}\cdot A_{j}= C_{ijk}\eta^{kl}A_{l}.
\end{equation}
Then we define quantum cohomology ring on $M$. In this ring we 
change $C_{ijk}$ into $C_{ijk}^{q}$ which we set to three point functions 
of topological sigma model (A-model) having as world sheet as $CP^{1}$.
 Metric is unchanged under the 
assumption of flat metric condition. Then multiplication 
rules are defined as,
\begin{eqnarray}
{\cal O}_{A_{i}}\cdot{\cal O}_{A_{j}}&=& C^{q}_{ijk}\eta^{kl}
{\cal O}_{A_{l}}\nonumber\\
   C^{q}_{ijk}&:=& \langle{\cal O}_{A_{i}}(z_{1}){\cal O}_{A_{j}}(z_{2})
{\cal O}_{A_{k}}(z_{k})
\rangle
\label{qc}
\end{eqnarray}
Since ${\cal M}_{0,0}^{M}$ is the moduli space of constant maps 
from $CP^{1}$ to $M$, it can be identified with $M$ itself.
Then  if we expand $C^{q}_{ijk}$ by $q$, we can see
 the coefficients of $q^{0}$ equals classical $C_{ijk}$.
In this sense,  this algebra is natural extension of the classical 
cohomology ring on $M$. We can see this algebra is commutative 
by definition. We assume it is also associative algebra. 
In the theory coupled with gravity, this condition is 
powerful to determine correlation functions in the small phase space.\cite{km}

Quantum cohomology ring is an effective notion in treating 
pure matter theory. Since in pure matter theory, 
$|{\cal O}_{A_{i}}\rangle$'s span entire Hilbert space, we can insert
identity operator 
$|{\cal O}_{A_{i}}\rangle\eta^{ij}\langle {\cal O}_{A_{j}}|$ into 
correlation functions. Especially fusion rules follow from this fact.
\begin{equation}
\langle {\cal O}_{A_{i}}{\cal O}_{A_{j}}*\rangle
= \langle {\cal O}_{A_{i}}{\cal O}_{A_{j}}{\cal O}_{A_{k}}\rangle
  \eta^{kl}\langle{\cal O}_{A_{l}}*\rangle 
\label{fus1}
\end{equation}
This relation can be rewritten using (\ref{qc}) as follows.
\begin{equation}
\langle {\cal O}_{A_{i}}{\cal O}_{A_{j}}*\rangle
= \langle ({\cal O}_{A_{i}}\cdot{\cal O}_{A_{j}})*\rangle 
\label{fus2}
\end{equation} 
Then by taking product of quantum cohomology ring successively,
we can reduce the number of inserted operators to two. In this 
way correlation functions of pure matter theory having world 
sheet as $CP^{1}$ are reduced 
to the problem of determination of all the three point functions.

We make one final remark of this section.
If this algebra has some relation ${\cal R}({\cal O}_{A_i})=0$, 
we can easily see from (\ref{fus2}),
\begin{equation}
\langle{\cal R}({\cal O}_{A_i})*\rangle = 0
\label{id}
\end{equation}
So we can compute correlation functions more effectively 
if we find non-trivial relation of the ring.   

\section{Strategy for Determination of Quantum Cohomology Ring of 
$M_{N}^{k}$}
In this paper, we treat the topological sigma model with a target space
which is the degree $k$ hypersurface $(k \leq N)$ in $CP^{N-1}$,
$M_{N}^{k}$.
\begin{eqnarray}
M_{N}^{k}:=\{(X_{1}:X_{2}:\cdots:X_{N})\in CP^{N-1}|
             X_{1}^{k}+\cdots+X_{N}^{k}=0\}
\label{a1}
\end{eqnarray}
 Since $M_{N}^{k}$ is a hypersurface of $CP^{N-1}$, we can choose a subring 
$H^{*}_{e}(M_{N}^{k})$ generated by Ka\"hler class 
$e \in H^{1,1}(M_{N}^{k})$.
Correspondingly, we assume that BRST-closed observables 
${\cal O}_{e^{\alpha}}\quad (\alpha = 0,1,\cdots,N-2)$ form a closed 
sub-algebra in quantum cohomology ring of $M_{N}^{k}$ (Operator algebra in 
pure matter theory). Then we investigate this sub-algebra 
$H^{*}_{q,e}(M_{N}^{k})$ in the following way. Operator product algebra is 
constructed by three point functions and metric as we mentioned 
in Section 2.
\begin{eqnarray}
\lefteqn{{\cal O}_{e^{\alpha}}\cdot {\cal O}_{e^{\beta}}
         = \langle {\cal O}_{e^{\alpha}}{\cal O}_{e^{\beta}}
           {\cal O}_{e^{\gamma}}\rangle \eta^{\gamma \delta}
           {\cal O}_{e^{\delta}}}\nonumber\\
&&\eta_{\gamma \delta}:= \langle {\cal O}_{e^{0}}
       {\cal O}_{e^{\gamma}}{\cal O}_{e^{\delta}}\rangle
       = \int_{M_{N}^{k}}e^{\gamma}\wedge e^{\delta}
       = k\delta_{\gamma+\delta,N-2}\nonumber\\
&&\eta_{\alpha \beta}\eta^{\beta\gamma}= \delta_{\alpha}^{\gamma}
\label{a2}
\end{eqnarray}
Correlation functions in pure matter theory satisfy the fusion 
rule.
\begin{eqnarray}
\langle {\cal O}_{e^{\alpha}}{\cal O}_{e^{\beta}}*\rangle
  = \langle {\cal O}_{e^{\alpha}}{\cal O}_{e^{\beta}}
           {\cal O}_{e^{\gamma}}\rangle \eta^{\gamma \delta}
          \langle {\cal O}_{e^{\delta}}*\rangle
\label{a3}
\end{eqnarray}
In (\ref{a2}) and (\ref{a3}) we assumed that insertion of 
${\cal O}_{w}\notin H^{*}_{q,e}(M_{N}^{k})$ into correlation
functions consisting of  elements of  
$H^{*}_{q,e}(M_{N}^{k})$ gives zero.
This assumption is justified in the case of $M_{N}^{N}$ in \cite{gmp}.
Other cases are discussed in section 1.
From the above definition we can easily see ${\cal O}_{e^{0}}$ acts 
trivially on $H^{*}_{q,e}(M_{N}^{k})$, and we regard ${\cal O}_{e^{0}}$
as identity. Three point functions are determined from the geometrical 
evaluation of correlation functions of topological sigma model
in previous section. 
\begin{eqnarray}
\lefteqn{\langle {\cal O}_{e^{\alpha}}(z_{1}){\cal O}_{e^{\beta}}(z_{2})
{\cal O}_{e^{\gamma}}(z_{3}) \rangle }\nonumber\\ 
&&= \sum_{d=0}^{\infty}\delta_{\alpha+\beta+\gamma,(N-k)d+N-2}
                 \int_{{\cal M}_{0,d}^{M_{N}^{k}}}
       c_{T}({\cal F})\wedge 
\phi_{1}^{*}(e^{\alpha})\wedge \phi_{2}^{*}(e^{\beta})
        \wedge \phi_{3}^{*}(e^{\gamma})\cdot q^{d}\nonumber\\
&& = \sum_{d=0}^{\infty}\delta_{\alpha+\beta+\gamma,(N-k)d+N-2} 
         \int_{{\cal M}_{0,d,3}^{M_{N}^{k}}}
        c_{T}({\cal F'})
\wedge\tilde\phi_{1}^{*}(e^{\alpha})\wedge \tilde\phi_{2}^{*}(e^{\beta})
        \wedge \tilde\phi_{3}^{*}(e^{\gamma})\cdot q^{d}
\label{def}
\end{eqnarray}
where
\begin{eqnarray}
\phi_{i}:{\cal M}_{0,d}^{M_{N}^{k}}&\mapsto& M_{N}^{k}\;\;
\phi_{i}(f)= f(z_{i})\nonumber\\ 
\tilde\phi_{i}:{\cal M}_{0,d,3}^{M_{N}^{k}}
&\mapsto& M_{N}^{k}\;\;
\tilde\phi_{i}(\{f,z_{1},z_{2},z_{3}\}/SL(2,C))= f(z_{i})
\label{evaluate}
\end{eqnarray}
${\cal M}_{0,d}^{M_{N}^{k}}$ and ${\cal M}_{0,d,3}^{M_{N}^{k}}$
 denote moduli spaces of holomorphic maps of degree $d$ 
from $CP^{1}$ to $M_{N}^{k}$ of 
pure matter theory and those of the theory coupled to gravity with three 
punctures
(operator insertion points) respectively. ${\cal F}$ and ${\cal F'}$ are 
obstruction
bundle coming from $H^{1}$.  We insert 
$\delta_{\alpha+\beta+\gamma,(N-k)d+N-2}$ 
to represent the topological selection rules explicitly. 
\begin{eqnarray}
&&\langle\prod_{j=1}^{m}{\cal O}_{e^{\alpha_{j}}}(z_{j})\rangle_{d}\neq 0
\nonumber\\
&\Longrightarrow& \sum_{j=1}^{m}\alpha_{j}
=\mbox{dim}_{C}(M_{N}^{k})+d\cdot c_{1}(T'M_{N}^{k})=
(N-2)+(N-k)d
\label{nksel}
\end{eqnarray}
$c_{1}(T'M_{N}^{k})$ is calculated from 
$c(T'M_{N}^{k})=j^{*}(c(T'CP^{N-1}))/j^{*}(c({\cal O}_{CP^{N-1}}(kH)))
=(1+e)^{N}/(1+ke)$
where $j$ represents natural embedding from $M_{N}^{k}$ to $CP^{N-1}$ and 
$e$ is K\"ahler form of $M_{N}^{k}$.

The equality between the first line and the second line of (\ref{def}) can 
be explained as follows. ${\cal M}_{0,d}^{M_{N}^{k}}$ has an internal 
$SL(2,C)$ which moves $\{f(z_{1}),f(z_{2}),f(z_{3})\}$
without changing the position of $f(CP^{1})$ in $M_{N}^{k}$.
In ${\cal M}_{0,d,3}^{M_{N}^{k}}$, these degrees of are 
killed by dividing by $SL(2,C)$ but the degrees of 
freedom that change the position of $\{z_{1},z_{2},z_{3}\}$
on $CP^{1}$ are added.
Since $SL(2,C)$
can be considered as the degrees of freedom which maps $\{0,1,\infty\}$
 to any distinct points $\{z_{1},z_{2},z_{3}\}$,
this difference cannot be distinguished under the action of 
the evaluation maps $\phi_{i},\tilde\phi_{i}$. As to equivalence of
$c_{T}({\cal F})$ and $c_{T}({\cal F'})$, we have to rely on numerical 
results. Equivalence in Calabi-Yau case was examined in \cite{tor} and
\cite{j}.

Then we determine 
$H^{*}_{e,q}(M_{N}^{k})$ with the following strategy.

1. Using the equality of (\ref{def}), we evaluate all the three point 
functions using the torus action method with the following
equation \cite{j}.
\begin{eqnarray}
\lefteqn{ \int_{{\cal M}_{0,d,3}^{M_{N}^{k}}}c_{T}({\cal F'})\wedge
        \tilde\phi_{1}^{*}(e^{\alpha})\wedge \tilde\phi_{2}^{*}(e^{\beta})
        \wedge \tilde\phi_{3}^{*}(e^{\gamma})}\nonumber\\
&&=\int_{{\cal M}_{0,d,3(gravity)}^{CP^{N-1}}}
        c_{T}({\tilde{\pi}}_{3}^{*}({\cal E}_{kd+1}))\wedge 
        \tilde\varphi_{1}^{*}(c_{1}^{\alpha}(H))\wedge 
        \tilde\varphi_{2}^{*}(c_{1}^{\beta}(H))\wedge 
        \tilde\varphi_{3}^{*}(c_{1}^{\gamma}(H))\nonumber\\
&&= \partial_{t_{\alpha}} \partial_{t_{\beta}} \partial_{t_{\gamma}} 
Res_{z}Res_{{h}}(\frac{1}{z}\log(\mbox{det}((g_{ij,j'i',d})
^{-1})\frac{1}{{h}}\int d\phi_{ij,d}\nonumber\\
&&\exp(-\frac{1}{2}\sum_{i,j,d}\frac{-d^{(N-2-(N-k)d)}(5^{i}z-5^{j}z)^{2}
\prod_{l=1}^{N}\prod_{a=1}^{d-1}(5^{i}az+5^{j}(d-a)z-
5^{l}dz)}{\prod_{a=1}^{kd-1}(5^{i}az+5^{j}(kd-a)z)}
\phi_{ij,d}\phi_{ji,d}\nonumber\\
&&+\sum_{i=1}^{N}
\frac{5^{i}kz}{\prod_{j\ne i}(5^{i}z-5^{j}z)}
\sum_{l=1}^{\infty}\frac{1}{l!}
\sum_{{d_{1},\cdots,d_{l},d_{*}\geq 1}\atop{j_{1},\cdots,j_{l},
j_{*}\ne i}}
(\frac{v_{ij_{1},d_{1}}}{z}+\cdots+\frac{v_{ij_{l},d_{l}}}{z})^{l-3}
\phi_{ij_{1},d_{1}}\cdots\phi_{ij_{l},d_{l}}\nonumber\\
&&\exp((5^{i}t_{1}z+\cdots+5^{i(N-2)}t_{N-2}z^{N-2})
(\frac{v_{ij_{1},d_{1}}}{z}+\cdots+\frac{v_{ij_{l},d_{l}}}{z})))))
|_{t_{*}=0}
\label{a4d}
\end{eqnarray}
where
\begin{eqnarray}
g_{ij,j'i',d}&:=&\frac{-d^{(N-2-(N-k)d)}(5^{i}z-5^{j}z)^{2}
\prod_{l=1}^{N}\prod_{a=1}^{d-1}(5^{i}az+5^{j}(d-a)z-
d5^{l}z)}{\prod_{a=1}^{kd-1}(5^{i}az+5^{j}(kd-a)z)}\nonumber\\
\\
v_{ij,d}&:=& \frac{d}{5^{i}-5^{j}}
\label{a4a}
\end{eqnarray}
\begin{eqnarray}
\tilde{\varphi}_{i}:{\cal M}_{0,d,3}^{CP^{N-1}}
&\mapsto &CP^{N-1},\;\;
\tilde{\varphi}(\{(z_{1},z_{2},z_{3}),f\}/SL(2,C)) 
= f(z_{i})\nonumber
\label{a4b}
\end{eqnarray}
where $H$ is hyperplane bundle on $CP^{N-1}$, $\tilde{\pi_{3}}$ is 
3-fold forgetful map from \\
${\cal M}_{0,d,3}^{CP^{N-1}}$ to 
${\cal M}_{0,d,0}^{CP^{N-1}}$ and 
${\cal E}_{kd+1}$ denotes direct image sheaf \\
$R_{\pi_{1}}^{0}(\tilde{\varphi}^{*}_{1}(kH))$ coming from forgetful map 
$\pi_{1}$ from 
${\cal M}_{0,d,1}^{CP^{N-1}}$ to 
${\cal M}_{0,d,0}^{CP^{N-1}}$.
 
2. We can consider ${\cal O}_{e}$ as the generator of 
$H^{*}_{q,e}(M_{N,k})$, and we only have to determine multiplication rule 
for ${\cal O}_{e}$. In other words, if we set
\begin{eqnarray}
F_{\alpha}^{N,k} : {\cal O}_{e}\cdot {\cal O}_{e^{\alpha}}
          & =& \sum_{d=0}^{[\frac{1+\alpha}{N-k}]}
            \langle {\cal O}_{e}{\cal O}_{e^{\alpha}}
             {\cal O}_{e^{N-3-\alpha+(N-k)d}}\rangle
             \frac{1}{k}{\cal O}_{e^{1+\alpha-(N-k)d}}\\
&&(k<N)\nonumber
\label{ad5}
\end{eqnarray}

\begin{eqnarray}
F_{\alpha}^{N,N} : {\cal O}_{e}\cdot {\cal O}_{e^{\alpha}}      
          & =& \langle {\cal O}_{e}{\cal O}_{e^{\alpha}}
             {\cal O}_{e^{N-3-\alpha}}\rangle
             \frac{1}{N}{\cal O}_{e^{1+\alpha}}\\
&&(k=N)\nonumber
\label{a5}
\end{eqnarray}
, $H_{q}^{*}(M_{N}^{k})$ is constructed under the assumption
of its commutativity and associativity as follows.
\begin{equation}
C[{\cal O}_{e},{\cal O}_{e^{2}},\cdots ,{\cal O}_{e^{N-2}}]/
 I[F^{N,k}_{1},F^{N,k}_{2},\cdots,F^{N,k}_{N-2}]
\label{a6}
\end{equation}
where $C[{\cal O}_{e},{\cal O}_{e^{2}},\cdots ,{\cal O}_{e^{N-2}}]$ 
denotes the polynomial ring generated by ${\cal O}_{e^{\alpha}}$ and 
$I[F^{N,k}_{1},F^{N,k}_{2},\cdots,F^{N,k}_{N-2}]$ is the ideal 
generated by 
$F^{N,k}_{\alpha}$'s. 

We calculate $F^{N,k}_{\alpha}$ for $k \leq N-2$ and
$N \leq 9$ case and find the ideal includes the following relation.
\begin{equation}
({\cal O}_{e})^{N-1}-k^{k}\cdot q\cdot ({\cal O}_{e})^{k-1}=0
\label{a7}
\end{equation}
Numerical results are shown in appendix A.
In this case, using (\ref{a7}) and \\
$\langle\prod_{i=1}^{N-2}
{\cal O}_{e}(z_{i})\rangle_{0}
 = k$, we can easily see  from (\ref{id})
\begin{equation}
\langle\prod_{i=1}^{N-2+(N-k)d}
{\cal O}_{e}(z_{i})\rangle = k^{kd+1}\cdot q^{d}.
\label{a8}
\end{equation}
\section{Reformulation as One Variable Polynomial Algebra}
With some algebra, we can rewrite the relations (\ref{a5}) into 
the form
\begin{eqnarray}
G_{\alpha}^{N,k}: {\cal O}_{e^{\alpha}}&=&
         ({\cal O}_{e})^{\alpha}-
       \sum_{d=1}^{[\frac{\alpha}{N-k}]}
       \gamma_{\alpha,d}^{N,k}({\cal O}_e)^{\alpha-(N-k)d}
       \cdot q^{d}\\
&&(2\leq \alpha \leq N-2)
\label{b1a}
\end{eqnarray}
\begin{eqnarray}
 G_{rel}^{N,k}:0 &=& ({\cal O}_{e})^{N-1}- 
      \sum_{d=1}^{[\frac{N-1}{N-k}]}
       \delta_{d}^{N,k}({\cal O}_{e})^{N-1-(N-k)d}
       \cdot q^{d}\\
&&(N>k)\nonumber
\label{b1b}
\end{eqnarray}
\begin{eqnarray}
 G_{\alpha}^{N,N}:{\cal O}_{e^{\alpha}}&=&
            (\prod_{j=1}^{\alpha}\gamma_{j}^{N,N}(q))
            ({\cal O}_{e})^{\alpha}\\
&&(2\leq \alpha \leq N-2)\nonumber
\label{b1c}
\end{eqnarray}
\begin{eqnarray}
 G_{rel}^{N,N}:0&=&(\prod_{j=1}^{N-2}\gamma_{j}^{N,N}(q))
({\cal O}_{e})^{N-1}\\
&&(N=k)\nonumber
\label{b1d}
\end{eqnarray}
where
\begin{equation}
\gamma_{j}^{N,N}(q):= N/\langle {\cal O}_{e}{\cal O}_{e^{j-1}}
             {\cal O}_{e^{N-2-j}}\rangle\nonumber\\
\label{b1e}
\end{equation}
Then we can realize $H^{*}_{q,e}(M^{k}_{N})$ as one variable 
polynomial algebra by regarding ${\cal O}_{e^{0}}$ as $1$, 
${\cal O}_{e}$ as $X$,
${\cal O}_{e^{\alpha}}$ as r.h.s of $G_{\alpha}^{N,k}$, 
and $G^{N,k}_{rel}$ 
as a relation.
 And if we set 
\begin{eqnarray}
 f^{N,k}_{0}(X)&:=&1,\;\;f^{N,k}_{1}(X):=X \nonumber\\
 f^{N,k}_{\alpha}(X)&:=&\mbox{r.h.s. of}\;\;G^{N,k}_{\alpha}  
\;\;\;\;(\alpha =2, \cdots, N-2)\nonumber\\
f^{N,k}_{rel.}(X)&:=& \mbox{r.h.s. of}\;\;G^{N,k}_{rel},
\label{polydef}
\end{eqnarray} 
 correlation functions are written in the residue 
form which follows from (\ref{a3}) as is well-known in \cite{fusion}, 
\begin{equation}                              
\langle {\cal O}_{e^{\alpha^{1}}}{\cal O}_{e^{\alpha^{2}}}
\cdots {\cal O}_{e^{\alpha^{l}}}\rangle_{M_{N}^{k}}
 = k\cdot\oint_{C_{\infty}}
(\frac{f^{N,k}_{\alpha_{1}}(X)\cdot f^{N,k}_{\alpha_{2}}(X)\cdots 
        f^{N,k}_{\alpha_{l}}(X)}{f^{N,k}_{rel}(X)})dX
\label{b2}
\end{equation}
where the integration contour  $C_{\infty}$ is a  small circle 
around $X=\infty$ in the counter-clockwise direction. 
These results are collected in Appendix B. 

{\bf Proof of (\ref{b2})}

First, we assume the following relation,
\begin{equation}
f^{N,k}_{\alpha_{1}}(X)\cdot f^{N,k}_{\alpha_{2}}(X)=h(X)f^{N,k}_{rel}(X)
+\langle
{\cal O}_{e^{\alpha_{1}}}{\cal O}_{e^{\alpha_{2}}}{\cal O}_{e^{h}}
\rangle\eta^{hm}f^{N,k}_{m}(X)
\label{prrel}
\end{equation}
where $h(X)$ is a certain polynomial of $X$. To be more precise, we 
have to ``check'' (\ref{prrel}) by numerical calculation to justify our 
construction but we heuristically rely on completeness of this algebra.
Then 
\begin{eqnarray}
\mbox{(r.h.s) of (\ref{b2})}&=&  k\cdot\oint_{C_{\infty}}
h(X)\cdot(f^{N,k}_{\alpha_{3}}(X)\cdot f^{N,k}_{\alpha_{4}}(X)\cdots 
        f^{N,k}_{\alpha_{l}}(X))dX \nonumber\\
&+&\langle
{\cal O}_{e^{\alpha_{1}}}{\cal O}_{e^{\alpha_{2}}}{\cal O}_{e^{h}}
\rangle\eta^{hm} k\cdot\oint_{C_{\infty}}
(\frac{f^{N,k}_{m}(X)\cdot f^{N,k}_{\alpha_{3}}(X)\cdots 
        f^{N,k}_{\alpha_{l}}(X)}{f^{N,k}_{rel}(X)})dX\nonumber\\
\label{prdec}
\end{eqnarray}
The first term of (\ref{prdec}) vanishes because if we set 
$X'= \frac{1}{X}$, we have terms with degree no more than $-2$ in the 
integrand.
And we have shown  fusion rule for representation of (\ref{b2}). 

Then by induction, we only have to show the following formula. 
\begin{equation}
k\cdot
\oint_{C_{\infty}}\frac{f^{N,k}_{\alpha_{1}}(X)\cdot 
f^{N,k}_{\alpha_{2}}(X)}{f^{N,k}_{rel}(X)}
 =k\cdot\delta_{\alpha_{1}+\alpha_{2},N-2}
\label{prtwo}
\end{equation}
But, by inserting ${\cal O}_{e^{0}}=1$ formally, (this insertion
does not change the value of correlation function) and using (\ref{prrel})
again, we only have to show the following relation instead of 
(\ref{prtwo}).
\begin{equation}
k\cdot
\oint_{C_{\infty}}\frac{f^{N,k}_{\alpha}(X)}{f^{N,k}_{rel}(X)}
 =k\cdot\delta_{\alpha,N-2}
\label{prone}
\end{equation}
(\ref{prone}) is trivial if we change the variable $X$ 
into $X'= \frac{1}{X}$.
\\
\\ 
 At first sight, this reformulation seems to be superficial, but 
we find some curious relation between $\gamma^{N,k}_{\alpha,d}$ for 
$k\leq N-2$ case.
\begin{eqnarray}
\mbox{relation 1}&&\nonumber\\
\gamma^{N,k}_{\alpha,1}&=& \gamma^{N-1,k}_{\alpha-1,1} 
 \qquad(k\leq N-2)\\
\mbox{relation 2}&&\nonumber\\
\gamma^{N,k}_{N-2,2}&=&\frac{(\gamma^{N,k}_{N-k,1})^{2}}{2}\\
&&((N-k)2=N-2)\nonumber\\
\gamma^{N,k}_{N-3,2}&=&\gamma^{N,k}_{N-k,1}
     (\frac{\gamma^{N,k}_{N-k+1,1}}{2}
      -\frac{\gamma^{N,k}_{N-k,1}}{4})\\
\gamma^{N,k}_{N-2,2}&=&\gamma^{N,k}_{N-k,1}
     (\frac{\gamma^{N,k}_{N-k+1,1}}{2}
      +\frac{\gamma^{N,k}_{N-k,1}}{4})\\
&&((N-k)2=N-3)\nonumber\\
\gamma^{N,k}_{N-4,2}&=&\gamma^{N,k}_{N-k,1}
     (\frac{\gamma^{N,k}_{N-k+2,1}}{2}
      -\frac{\gamma^{N,k}_{N-k+1,1}}{4}
     -\frac{\gamma^{N,k}_{N-k,1}}{8})\\     
\gamma^{N,k}_{N-3,2}&=&\frac{(\gamma^{N,k}_{N-k+1,1})^{2}}{2}\\
\gamma^{N,k}_{N-2,2}&=&\gamma^{N,k}_{N-k,1}
     (\frac{\gamma^{N,k}_{N-k+2,1}}{2}
      +\frac{\gamma^{N,k}_{N-k+1,1}}{4}
     +\frac{\gamma^{N,k}_{N-k,1}}{8})\\ 
&&((N-k)2=N-4)\nonumber\\
\gamma^{N,k}_{N-5,2}&=&\gamma^{N,k}_{N-k,1}
     (\frac{\gamma^{N,k}_{N-k+3,1}}{2}
      -\frac{\gamma^{N,k}_{N-k+2,1}}{4}
     -\frac{\gamma^{N,k}_{N-k+1,1}}{8}
     -\frac{\gamma^{N,k}_{N-k,1}}{16})\\
\gamma^{N,k}_{N-4,2}&=&\gamma^{N,k}_{N-k+1,1}
     (\frac{\gamma^{N,k}_{N-k+2,1}}{2}
      -\frac{\gamma^{N,k}_{N-k+1,1}}{4})\nonumber\\
&&+\frac{\gamma^{N,k}_{N-k,1}}{2}
     (\frac{\gamma^{N,k}_{N-k+2,1}}{2}
      -\frac{\gamma^{N,k}_{N-k+1,1}}{4}
     -\frac{\gamma^{N,k}_{N-k,1}}{8})\\
\gamma^{N,k}_{N-3,2}&=&\gamma^{N,k}_{N-k+1,1}
     (\frac{\gamma^{N,k}_{N-k+2,1}}{2}
      +\frac{\gamma^{N,k}_{N-k+1,1}}{4})\nonumber\\
&&-\frac{\gamma^{N,k}_{N-k,1}}{2}
     (\frac{\gamma^{N,k}_{N-k+2,1}}{2}
      -\frac{\gamma^{N,k}_{N-k+1,1}}{4}
     -\frac{\gamma^{N,k}_{N-k,1}}{8})\\
\gamma^{N,k}_{N-2,2}&=&\gamma^{N,k}_{N-k,1}
     (\frac{\gamma^{N,k}_{N-k+3,1}}{2}
      +\frac{\gamma^{N,k}_{N-k+2,1}}{4}
     +\frac{\gamma^{N,k}_{N-k+1,1}}{8}
     +\frac{\gamma^{N,k}_{N-k,1}}{16})\\
&&((N-k)2=N-5)\nonumber\\
\mbox{relation 3}&&\nonumber\\
\gamma^{N,k}_{N-2,3}&=&\gamma^{N,k}_{N-k,1}\gamma^{N,k}_{2(N-k),2}\\
&& ((N-k)3=N-2)\nonumber
\label{b3c}
\end{eqnarray} 
We can reconstruct some of the above relations from the compatibility
of the expansion form of (37) and relation (\ref{a7}), but 
we are not sure that all of them follow from it at this stage.
With these relations, we can figure out some characteristic feature 
of $H^{*}_{q,e}(M_{N}^{k})$.\\
First, quantum correction of degree 1 to $H^{*}_{q,e}(M_{N}^{k})$
does not depend on $N$, which can be easily seen from relation 1.
So we think these correction coefficients $\gamma^{k}_{\alpha,1}
(:=\gamma^{N,k}_{N-k+\alpha-1,1})$ play central role in the ring
when $k\leq N-2$. In other words, we expect all the higher degree
quantum correction coefficients are determined by 
$\gamma^{k}_{\alpha,1}$. Relation 2 are found from these speculations.
Second, from the expansion form of (\ref{b1a}), degree d coefficients
of ${\cal O}_{e^{\alpha}}$ occur when $\alpha\geq (N-k)d $ holds.
Then if $k\leq [\frac{N}{2}]+1$, no corrections occur from sectors 
with  degree greater than 1. But degree 1 corrections remain stable 
since they exist as long as $\alpha$ is no less than $N-k$.
This seems to support our first speculation. We will show some 
examples of these features using the results of 
$H^{*}_{q,e}(M_{N}^{6})$.
\begin{eqnarray}
H^{*}_{q,e}(M_{8}^{6})&&\nonumber\\
{\cal O}_{e}&=& X\nonumber\\
{\cal O}_{e^{2}}&=& X^{2}-\gamma^{6}_{1,1}q\nonumber\\
{\cal O}_{e^{3}}&=& X^{3}-\gamma^{6}_{2,1}Xq\nonumber\\
{\cal O}_{e^{4}}&=& X^{4}-\gamma^{6}_{3,1}X^{2}q
            -\gamma^{6}_{1,1}
     (\frac{\gamma^{6}_{3,1}}{2}
      -\frac{\gamma^{6}_{2,1}}{4}
     -\frac{\gamma^{6}_{1,1}}{8})q^{2}\nonumber\\
{\cal O}_{e^{5}}&=& X^{5}-\gamma^{6}_{4,1}X^{3}q
              - \frac{(\gamma^{6}_{2,1})^{2}}{2}Xq^{2}\nonumber\\
{\cal O}_{e^{6}}&=& X^{6}-\gamma^{6}_{5,1}X^{4}q
            -\gamma^{6}_{1,1}
     (\frac{\gamma^{6}_{3,1}}{2}
      +\frac{\gamma^{6}_{2,1}}{4}
     +\frac{\gamma^{6}_{1,1}}{8})X^{2}q^{2}\nonumber\\
&&-(\gamma^{6}_{1,1})^{2}(\frac{\gamma^{6}_{3,1}}{2}
      -\frac{\gamma^{6}_{2,1}}{4}
     -\frac{\gamma^{6}_{1,1}}{8})q^{3}\\
H^{*}_{q,e}(M_{9}^{6})&&\nonumber\\
{\cal O}_{e}&=& X\nonumber\\
{\cal O}_{e^{2}}&=& X^{2}\nonumber\\
{\cal O}_{e^{3}}&=& X^{3}-\gamma^{6}_{1,1}q\nonumber\\
{\cal O}_{e^{4}}&=& X^{4}-\gamma^{6}_{2,1}X^{1}q\nonumber\\
{\cal O}_{e^{5}}&=& X^{5}-\gamma^{6}_{3,1}X^{2}q\nonumber\\
{\cal O}_{e^{6}}&=& X^{6}-\gamma^{6}_{4,1}X^{3}q
              - \gamma^{6}_{1,1}
     (\frac{\gamma^{6}_{2,1}}{2}
      -\frac{\gamma^{6}_{1,1}}{4})q^{2}\nonumber\\
{\cal O}_{e^{7}}&=& X^{7}-\gamma^{6}_{5,1}X^{4}q
            -\gamma^{6}_{1,1}
     (\frac{\gamma^{6}_{2,1}}{2}
      +\frac{\gamma^{6}_{1,1}}{4})Xq^{2}  \\
H^{*}_{q,e}(M_{10}^{6})&&\nonumber\\
{\cal O}_{e}&=&X \nonumber\\
{\cal O}_{e^{2}}&=&X^{2}\nonumber\\
{\cal O}_{e^{3}}&=&X^{3}\nonumber\\
{\cal O}_{e^{4}}&=& X^{4}-\gamma^{6}_{1,1}q\nonumber\\
{\cal O}_{e^{5}}&=& X^{5}-\gamma^{6}_{2,1}Xq\\
{\cal O}_{e^{6}}&=& X^{6}-\gamma^{6}_{3,1}X^{2}q\nonumber\\
{\cal O}_{e^{7}}&=& X^{7}-\gamma^{6}_{4,1}X^{3}q\nonumber\\
{\cal O}_{e^{8}}&=& X^{8}-\gamma^{6}_{5,1}X^{4}q
-\frac{(\gamma_{1,1}^{6})^{2}}{2}q^{2}\\
H^{*}_{q,e}(M_{N}^{6})&&\qquad(N\geq 11)\nonumber\\
{\cal O}_{e^{k}}&=& X^{k}\qquad(1\leq k\leq N-7)\nonumber\\
{\cal O}_{e^{N-7+\alpha}}&=& X^{N-7+\alpha}-
         \gamma^{6}_{\alpha,1}X^{\alpha-1}q
    (1\leq \alpha\leq 5)
\label{b4a}
\end{eqnarray}
where
\begin{equation}
\gamma^{6}_{1,1}=720,\gamma^{6}_{2,1}=6984,\gamma^{6}_{3,1}=23328,
\gamma^{6}_{4,1}=39672,\gamma^{6}_{5,1}=45936\nonumber      
\label{b4b}
\end{equation}

\section{Geometrical Interpretation}
In this section, we will briefly discuss why relation (\ref{a7}) or 
equation (\ref{a8}) holds from the geometrical point of view. Since 
$M_{N}^{k}$ is hypersurface in $CP^{N-1}$, ${\cal M}^{d}_{N,k,(matter)}$
is submanifold of ${\cal M}^{d}_{CP^{N-1},(matter)}$. So we can 
expect the following formula.
\begin{equation}
\langle \prod_{i=1}^{N-2+(N-k)d}{\cal O}_{e}(z_{i})
         \rangle = \int_{{\cal M}_{0,d}^{M_{N}^{k}}}
         \bigwedge_{i=1}^{N-2+(N-k)d}\phi^{*}_{i}(e)  
=\int_{{\cal M}^{CP^{N-1}}_{0,d}}c_{(N,k)}\wedge
         (\bigwedge_{i=1}^{N-2+(N-k)d}\varphi^{*}_{i}(e))
\label{c1a}
\end{equation}
where
\begin{eqnarray}
\phi_{i}:{\cal M}_{0,d}^{M_{N}^{k}}&\mapsto& M_{N}^{k},\;\;
\phi_{i}(f)= f(z_{i})\nonumber\\
\varphi_{i}:{\cal M}_{0,d}^{CP^{N-1}}&\mapsto& CP^{N-1},\;\;
\varphi_{i}(f)= f(z_{i})
\end{eqnarray}
$c_{(N,k)}$ is the form which impose the following condition on 
$f\in {\cal M}^{CP^{N-1}}_{0,d}$.
\begin{equation}
f(CP^{1})\subset M_{N}^{k}
\label{c2}
\end{equation}
Since holomorphic map from $CP^{1}$ to $CP^{N-1}$ of degree d 
is described by the polynomial map as follows, we can roughly 
compactify ${\cal M}^{CP^{N-1}}_{0,d}$ into 
$CP^{N(d+1)-1}$.
\begin{eqnarray}
\lefteqn{(s:t)\mapsto (\sum_{j=0}^{d}a^{1}_{j}s^{j}t^{d-j}:
                       \sum_{j=0}^{d}a^{2}_{j}s^{j}t^{d-j}:
                       \cdots :
                       \sum_{j=0}^{d}a^{N}_{j}s^{j}t^{d-j})}\nonumber\\
&&:=(A^{1}_{d}(s,t):A^{2}_{d}(s,t):\cdots:A^{N}_{d}(s,t))\\
&&{\overline{\cal M}}^{CP^{N-1}}_{0,d}=(a^{1}_{0}:a^{1}_{0}:
\cdots:a^{1}_{d}:\cdots:\cdots:a^{N}_{0}:\cdots:a^{N}_{d})
\label{c3}
\end{eqnarray}
Using this compactification, we can realize $c_{(N,k)}$ as follows.
\begin{eqnarray}
\lefteqn{f(CP^{1})\subset M_{N}^{k}}\nonumber\\
&&\Longleftrightarrow \sum_{i=1}^{N}
(\sum_{j=0}^{d}a^{i}_{j}s^{j}t^{d-j})^{k}=0
\quad\mbox{for all}\quad(s,t)\nonumber\\
&&\Longleftrightarrow \sum_{l=1}^{kd}f_{l}(a_{j}^{i})s^{kd-l}t^{l}
=0\quad\mbox{for all}\quad(s,t)\nonumber\\ 
&&\Longleftrightarrow f_{l}(a_{j}^{i})= 0\qquad l=0,1,\cdots,kd+1
\label{c4}
\end{eqnarray}
Since each $f_{l}(a_{j}^{i})$ is a homogeneous polynomial of 
$a_{j}^{i}$ of degree $k$, we can regard them as 
$k\tilde{e}:=k c_{1}(\tilde{H})$ where $\tilde{H}$ is hyperplane 
bundle of $CP^{N(d+1)-1}$. And we have   
\begin{equation}
c_{(N,k)}= (k\tilde{e})^{kd+1}
\label{c5}
\end{equation} 
We can easily see that $\varphi^{*}_{i}(e)=\tilde{e}$ from the 
definition of $\varphi_{i}$ and we have
\begin{eqnarray}
\langle\prod_{i=1}^{N-2+(N-k)d}
{\cal O}_{e}(z_{i})\rangle_{c}&=&\int_{CP^{N(d+1)-1}}(k\tilde{e})^{kd+1}
\wedge(\tilde{e})^{N-2+(N-k)d}\nonumber\\
&=&k^{kd+1}
\label{c6}
\end{eqnarray}
Our calculation in section 3 tells us this compactification gives an  
exact result when $k\leq N-2$. To derive this bound geometrically,
we have to analyze points in $CP^{N(d+1)-1}$ added in the process 
of compactification. As was said in \cite{mnmj} and \cite{mp},
these points are characterized by the fact that all $A_{d}^{i}$'s 
have common divisor. This situation can be described by the 
following sequence of maps.
\begin{eqnarray}
\lefteqn{CP^{N(d+1)-1}\stackrel{\eta_{1}}{\longleftarrow}}
\nonumber\\
&&CP^{N(d)-1}\times CP^{1}\stackrel{\eta_{2}}{\longleftarrow}
CP^{N(d-1)-1}\times (CP^{1})^{2}\stackrel{\eta_{3}}{\longleftarrow}
\nonumber\\
&&\cdots\nonumber\\
&&CP^{N(d-k+1)-1}\times (CP^{1})^{k}
\stackrel{\eta_{k+1}}{\longleftarrow}
CP^{N(d-k)-1}\times (CP^{1})^{k+1}
\stackrel{\eta_{k+2}}{\longleftarrow}
\nonumber\\  
&&\cdots\nonumber\\
&&CP^{2N-1}\times (CP^{1})^{d-1}
\stackrel{\eta_{d}}{\longleftarrow}
CP^{N-1}\times (CP^{1})^{d}
\label{c7a}
\end{eqnarray}
where
\begin{eqnarray}
\lefteqn{\eta_{j}: CP^{N(d-j+1)-1}\times (CP^{1})^{j}
\mapsto CP^{N(d-j+2)-1}\times (CP^{1})^{j-1}}\nonumber\\
&&((A_{d-j}^{1}(s,t),\cdots,A_{d-j}^{N}(s,t)),(a^{1}s+b^{1}t),
\cdots,(a^{j}s+b^{j}t))\nonumber\\
&&\mapsto
((A_{d-j}^{1}(s,t)(a^{1}s+b^{1}t),\cdots,
A_{d-j}^{N}(s,t)(a^{1}s+b^{1}t))\nonumber\\
&&,(a^{2}s+b^{2}t),\cdots,(a^{j}s+b^{j}t))
\label{c7b}
\end{eqnarray}
 In the  calculation, we must treat $Im(\eta_{j})$
carefully. Then why (\ref{c6}) is correct in $k\leq N-2$ case ?
This can be understood as follows. Consider the first non-trivial 
boundary $Im(\eta_{1})$.
\begin{equation}
\eta_{1}(CP^{Nd-1}\times CP^{1})
= ((as+bt)A_{d-1}^{1}(s,t),\cdots,(as+bt)A_{d-1}^{N}(s,t))
\label{c8}
\end{equation}
For these points, the condition $f(CP^{1})\subset M_{N}^{k}$
acts only on $A_{d-1}^{i}(s,t)$'s and the remaining degrees of 
freedom come from $CP^{1}$ and constrained $A_{d-1}^{i}(s,t)$'s, i.e.
\begin{eqnarray}
{}^{\sharp}(\mbox{degrees of freedom})&=& 1+Nd-1-(k(d-1)+1)\nonumber\\
    &=& (N-k)d+k-1
\label{c9}
\end{eqnarray}                         
Then if the condition $k\leq N-2$ is satisfied, we have 
\begin{eqnarray}
(N-k)d+k-1 &<& (N-k)d + N-2\nonumber\\
           &=& \mbox{dim}({\cal M}^{M_{N}^{k}}_{0,d})
\label{c10}
\end{eqnarray}
(\ref{c9}) tells us that the condition (\ref{c4}) may make 
the contribution from $Im(\eta_{1})$ be a space whose
dimension is no less than $\mbox{dim}
({\cal M}^{M_{N}^{k}}_{0,d})$.
But in $k\leq N-2$ case, (\ref{c10}) assures us that it is 
irrelevant in view of the  dimensional counting.     
\section{Conclusion}
 Our main result of this paper is the determination 
of the bound $k\leq N-2$. Under this bound, principal relation 
of quantum cohomology ring is written in a simple form, 
$({\cal O}_{e})^{N-1}= k^{k}q({\cal O}_{e})^{k-1}$, which is a 
natural generalization of the well-known result of 
$CP^{N-2}$ model, $({\cal O}_{e})^{N-1}= q$ \cite{va}.
The ring $H^{*}_{q,e}(M_{N}^{k})$ is mainly characterized 
by $k$, so polynomial representations of operators with 
different $N$ are alike with each other. These seem to be 
determined by the correction coefficients 
$\gamma^{k}_{\alpha,1}$ coming from  holomorphic maps of 
degree 1  which are invariant under variation of $N$, though 
we cannot give complete formulation in this paper.\\
 We give geometrical interpretation of this bound in 
section 5 but this argument does not explain why insertion 
of operator ${\cal O}_{e^{2}}$ and $({\cal O}_{e})^{2}$
are distinct even if $k\leq N-2$. These insertions cannot 
be distinguished by our simple logic. Looking around the 
situations, it seems to be  effective only with the insertion
of BRST-closed operator induced from K\"ahler forms  
in treating hypersurface in weighted projective space.
Of course, as can be seen in \cite{bert}, moduli spaces 
of manifolds like Grassmannians which are defined 
as homogeneous spaces  are compactified without ambiguous 
process like the one in (\ref{c4}). In this case,
 such troubles do not 
arise.\\
Finally, we discuss what our results tells us with respect 
to $k= N-1,N$ case. At least, it supports our assertion 
that the first term of $N$-expansion of $\langle
\prod_{i=1}^{N-2}
{\cal O}_{e}(z_{i})\rangle$   on Calabi-Yau manifold 
$M_{N}^{N}$ comes from the compactifiaction treated in 
this paper but explanation of correction terms from 
this point of view is still not clear.

{\bf Acknowledgment}
\smallskip
I'd like to thank Dr.K.Hori and Prof.T.Eguchi
for many useful discussions. I also 
thank 
Dr.Y.Sun and Dr.M.Nagura for kind encouragement.
\newpage

\section*{\bf Appendix A  Multiplication Rules of $H_{q,e}^{*}(M_{N}^{k})$}
\begin{eqnarray}
\lefteqn{H_{q,e}^{*}(M_{N}^{1})}\nonumber\\
&&{\cal O}_{e}\cdot{\cal O}_{e^{\alpha}}=
{\cal O}_{e^{\alpha+1}}\quad(0\leq\alpha\leq N-3)\quad
{\cal O}_{e}\cdot{\cal O}_{e^{N-2}}=q
\label{d1}
\end{eqnarray}
\begin{eqnarray}
\lefteqn{H_{q,e}^{*}(M_{N}^{2})}\nonumber\\
&&{\cal O}_{e}\cdot{\cal O}_{e^{\alpha}}=
{\cal O}_{e^{\alpha+1}}\quad(0\leq\alpha\leq N-4)\quad
{\cal O}_{e}\cdot{\cal O}_{e^{N-3}}={\cal O}_{e^{N-2}}+2q\nonumber\\
&&{\cal O}_{e}\cdot{\cal O}_{e^{N-2}}= 2{\cal O}_{e}q
\label{d2}
\end{eqnarray}
\begin{eqnarray}
\lefteqn{H_{q,e}^{*}(M_{5}^{3})}\nonumber\\
&&{\cal O}_{e}\cdot{\cal O}_{e}={\cal O}_{e^{2}}+6q\quad
{\cal O}_{e}\cdot{\cal O}_{e^{2}}={\cal O}_{e^{3}}+
15{\cal O}_{e} q\nonumber\\
&&{\cal O}_{e}\cdot{\cal O}_{e^{3}}= 6{\cal O}_{e^{2}}q
+36q^{2}
\label{d3}
\end{eqnarray}
\begin{eqnarray}
\lefteqn{H_{q,e}^{*}(M_{N}^{3})\quad(N\geq 6)}\nonumber\\
&&{\cal O}_{e}\cdot{\cal O}_{e^{\alpha}}=
{\cal O}_{e^{\alpha+1}}\quad(0\leq\alpha\leq N-5)\nonumber\\
&&{\cal O}_{e}\cdot{\cal O}_{e^{N-4}}=
{\cal O}_{e^{N-3}}+6q\quad
{\cal O}_{e}\cdot{\cal O}_{e^{N-3}}={\cal O}_{e^{N-2}}+
15{\cal O}_{e} q\nonumber\\
&&{\cal O}_{e}\cdot{\cal O}_{e^{N-2}}= 6{\cal O}_{e^{2}}q
\label{d4}
\end{eqnarray}
\begin{eqnarray}
\lefteqn{H_{q,e}^{*}(M_{6}^{4})}\nonumber\\
&&{\cal O}_{e}\cdot{\cal O}_{e}={\cal O}_{e^{2}}+24q\nonumber\\
&&{\cal O}_{e}\cdot{\cal O}_{e^{2}}={\cal O}_{e^{3}}
+104{\cal O}_{e}q\nonumber\\
&&{\cal O}_{e}\cdot{\cal O}_{e^{3}}={\cal O}_{e^{4}}
+104{\cal O}_{e^{2}}q+2784q^{2}\nonumber\\
&&{\cal O}_{e}\cdot{\cal O}_{e^{4}}=24{\cal O}_{e^{3}}q
+2784{\cal O}_{e}q^{2}
\label{d4a}
\end{eqnarray}
\begin{eqnarray}
\lefteqn{H_{q,e}^{*}(M_{7}^{4})}\nonumber\\
&&{\cal O}_{e}\cdot{\cal O}_{e}={\cal O}_{e^{2}}\nonumber\\
&&{\cal O}_{e}\cdot{\cal O}_{e^{2}}={\cal O}_{e^{3}}+24q\nonumber\\
&&{\cal O}_{e}\cdot{\cal O}_{e^{3}}={\cal O}_{e^{4}}
+104{\cal O}_{e}q\nonumber\\
&&{\cal O}_{e}\cdot{\cal O}_{e^{4}}={\cal O}_{e^{5}}
+104{\cal O}_{e^{2}}q\nonumber\\
&&{\cal O}_{e}\cdot{\cal O}_{e^{5}}=24{\cal O}_{e^{3}}q
+576q^{2}
\label{d5}
\end{eqnarray} 
\begin{eqnarray}
\lefteqn{H_{q,e}^{*}(M_{N}^{4})}\nonumber\\
&&{\cal O}_{e}\cdot{\cal O}_{e^{\alpha}}=
{\cal O}_{e^{\alpha+1}}\quad
(0\leq\alpha\leq N-6)\nonumber\\
&&{\cal O}_{e}\cdot{\cal O}_{e^{N-5}}=
{\cal O}_{e^{N-4}}+24q\nonumber\\
&&{\cal O}_{e}\cdot{\cal O}_{e^{N-4}}={\cal O}_{e^{N-3}}
+104{\cal O}_{e}q\nonumber\\
&&{\cal O}_{e}\cdot{\cal O}_{e^{N-3}}={\cal O}_{e^{N-2}}
+104{\cal O}_{e^{2}}q\nonumber\\
&&{\cal O}_{e}\cdot{\cal O}_{e^{N-2}}=24{\cal O}_{e^{3}}q
\label{d6}
\end{eqnarray}
\begin{eqnarray}
\lefteqn{H_{q,e}^{*}(M_{7}^{5})}\nonumber\\
&&{\cal O}_{e}\cdot{\cal O}_{e}={\cal O}_{e^{2}}+120q\nonumber\\
&&{\cal O}_{e}\cdot{\cal O}_{e^{2}}={\cal O}_{e^{3}}
+770{\cal O}_{e}q\nonumber\\
&&{\cal O}_{e}\cdot{\cal O}_{e^{3}}={\cal O}_{e^{4}}
+1345{\cal O}_{e^{2}}q+211200q^{2}\nonumber\\
&&{\cal O}_{e}\cdot{\cal O}_{e^{4}}= {\cal O}_{e^{5}}
+770{\cal O}_{e^{3}}q
+692500{\cal O}_{e}q^{2}\nonumber\\
&&{\cal O}_{e}\cdot{\cal O}_{e^{5}}=
120{\cal O}_{e^{4}}q
+211200{\cal O}_{e^{2}}q^{2}
\label{d7}
\end{eqnarray}
\begin{eqnarray}
\lefteqn{H_{q,e}^{*}(M_{8}^{5})}\nonumber\\
&&{\cal O}_{e}\cdot{\cal O}_{e}={\cal O}_{e^{2}}\nonumber\\
&&{\cal O}_{e}\cdot{\cal O}_{e^{2}}={\cal O}_{e^{3}}+120q\nonumber\\
&&{\cal O}_{e}\cdot{\cal O}_{e^{3}}={\cal O}_{e^{4}}
+770{\cal O}_{e}q\nonumber\\
&&{\cal O}_{e}\cdot{\cal O}_{e^{4}}={\cal O}_{e^{5}}
+1345{\cal O}_{e^{2}}q\nonumber\\
&&{\cal O}_{e}\cdot{\cal O}_{e^{5}}= {\cal O}_{e^{6}}
+770{\cal O}_{e^{3}}q
+99600q^{2}\nonumber\\
&&{\cal O}_{e}\cdot{\cal O}_{e^{6}}=
120{\cal O}_{e^{4}}q
+99600{\cal O}_{e}q^{2}
\label{d7a}
\end{eqnarray}
\begin{eqnarray}
\lefteqn{H_{q,e}^{*}(M_{9}^{5})}\nonumber\\
&&{\cal O}_{e}\cdot{\cal O}_{e}={\cal O}_{e^{2}}\nonumber\\
&&{\cal O}_{e}\cdot{\cal O}_{e^{2}}={\cal O}_{e^{3}}\nonumber\\
&&{\cal O}_{e}\cdot{\cal O}_{e^{3}}={\cal O}_{e^{4}}+120q\nonumber\\
&&{\cal O}_{e}\cdot{\cal O}_{e^{4}}={\cal O}_{e^{5}}
+770{\cal O}_{e}q\nonumber\\
&&{\cal O}_{e}\cdot{\cal O}_{e^{5}}={\cal O}_{e^{6}}
+1345{\cal O}_{e^{2}}q\nonumber\\
&&{\cal O}_{e}\cdot{\cal O}_{e^{6}}= {\cal O}_{e^{7}}
+770{\cal O}_{e^{3}}q\nonumber\\
&&{\cal O}_{e}\cdot{\cal O}_{e^{7}}=
+120{\cal O}_{e^{4}}q
+14400q^{2}
\label{d8}
\end{eqnarray}
\begin{eqnarray}
H_{q,e}^{*}(M_{N}^{5})\quad(N\geq 10)&&\nonumber\\
{\cal O}_{e}\cdot{\cal O}_{e^{\alpha}}
&=&{\cal O}_{e^{\alpha+1}}\quad(0\leq\alpha\leq N-7)\nonumber\\
{\cal O}_{e}\cdot{\cal O}_{e^{N-6}}&=&{\cal O}_{e^{N-5}}+120q\nonumber\\
{\cal O}_{e}\cdot{\cal O}_{e^{N-5}}&=&{\cal O}_{e^{N-4}}
+770{\cal O}_{e}q\nonumber\\
{\cal O}_{e}\cdot{\cal O}_{e^{N-4}}&=&{\cal O}_{e^{N-3}}
+1345{\cal O}_{e^{2}}q\nonumber\\
{\cal O}_{e}\cdot{\cal O}_{e^{N-3}}&=& {\cal O}_{e^{N-2}}
+770{\cal O}_{e^{3}}q\nonumber\\
{\cal O}_{e}\cdot{\cal O}_{e^{N-2}}&=&
120{\cal O}_{e^{4}}q
\label{d9}
\end{eqnarray}
\newpage
\begin{eqnarray}
H_{q,e}^{*}(M_{8}^{6})&&\nonumber\\
{\cal O}_{e}\cdot{\cal O}_{e}&=&{\cal O}_{e^{2}}+720q\nonumber\\
{\cal O}_{e}\cdot{\cal O}_{e^{2}}&=&{\cal O}_{e^{3}}
+6264{\cal O}_{e}q\nonumber\\
{\cal O}_{e}\cdot{\cal O}_{e^{3}}&=&{\cal O}_{e^{4}}
+16344{\cal O}_{e^{2}}q+18843840q^{2}\nonumber\\
{\cal O}_{e}\cdot{\cal O}_{e^{4}}&=& {\cal O}_{e^{5}}
+16344{\cal O}_{e^{3}}q
+131458464{\cal O}_{e}q^{2}\nonumber\\
{\cal O}_{e}\cdot{\cal O}_{e^{5}}&=&
{\cal O}_{e^{6}}
+6264{\cal O}_{e^{4}}q
+131458464{\cal O}_{e^{2}}q^{2}\nonumber\\
&&+144069995520q^{3}\nonumber\\
{\cal O}_{e}\cdot{\cal O}_{e^{6}}&=&
+720{\cal O}_{e^{5}}q
+18843840{\cal O}_{e^{3}}q^{2}\nonumber\\
&&+144069995520{\cal O}_{e}q^{3}
\label{d10}
\end{eqnarray}
\begin{eqnarray}
H_{q,e}^{*}(M_{9}^{6})&&\nonumber\\
{\cal O}_{e}\cdot{\cal O}_{e}&=&{\cal O}_{e^{2}}\nonumber\\
{\cal O}_{e}\cdot{\cal O}_{e^{2}}&=&{\cal O}_{e^{3}}+720q\nonumber\\
{\cal O}_{e}\cdot{\cal O}_{e^{3}}&=&{\cal O}_{e^{4}}
+6264{\cal O}_{e}q\nonumber\\
{\cal O}_{e}\cdot{\cal O}_{e^{4}}&=&{\cal O}_{e^{5}}
+16344{\cal O}_{e^{2}}q\nonumber\\
{\cal O}_{e}\cdot{\cal O}_{e^{5}}&=& {\cal O}_{e^{6}}
+16344{\cal O}_{e^{3}}q
+14152320q^{2}\nonumber\\
{\cal O}_{e}\cdot{\cal O}_{e^{6}}&=&
{\cal O}_{e^{7}}
+6264{\cal O}_{e^{4}}q
+44006976{\cal O}_{e}q^{2}\nonumber\\
{\cal O}_{e}\cdot{\cal O}_{e^{7}}&=&
720{\cal O}_{e^{5}}q
+14152320{\cal O}_{e^{2}}q^{2}
\label{d11}
\end{eqnarray}
\begin{eqnarray}
H_{q,e}^{*}(M_{10}^{6})&&\nonumber\\
{\cal O}_{e}\cdot{\cal O}_{e}&=&{\cal O}_{e^{2}}\nonumber\\
{\cal O}_{e}\cdot{\cal O}_{e^{2}}&=&{\cal O}_{e^{3}}\nonumber\\
{\cal O}_{e}\cdot{\cal O}_{e^{3}}&=&{\cal O}_{e^{4}}
+720q\nonumber\\
{\cal O}_{e}\cdot{\cal O}_{e^{4}}&=&{\cal O}_{e^{5}}
+6264{\cal O}_{e}q\nonumber\\
{\cal O}_{e}\cdot{\cal O}_{e^{5}}&=&{\cal O}_{e^{6}}
+16344{\cal O}_{e^{2}}q\nonumber\\
{\cal O}_{e}\cdot{\cal O}_{e^{6}}&=& {\cal O}_{e^{6}}
+16344{\cal O}_{e^{3}}q\nonumber\\
{\cal O}_{e}\cdot{\cal O}_{e^{7}}&=&
{\cal O}_{e^{8}}
+6264{\cal O}_{e^{4}}q
+4769280{\cal O}_{e}q^{2}\nonumber\\
{\cal O}_{e}\cdot{\cal O}_{e^{8}}&=&
720{\cal O}_{e^{5}}q
+4769280{\cal O}_{e^{2}}q^{2}
\label{d12}
\end{eqnarray}
\newpage
\begin{eqnarray}
H_{q,e}^{*}(M_{9}^{7})&&\nonumber\\
{\cal O}_{e}\cdot{\cal O}_{e}&=&{\cal O}_{e^{2}}+5040q\nonumber\\
{\cal O}_{e}\cdot{\cal O}_{e^{2}}&=&{\cal O}_{e^{3}}
+56196{\cal O}_{e}q\nonumber\\
{\cal O}_{e}\cdot{\cal O}_{e^{3}}&=&{\cal O}_{e^{4}}
+200452{\cal O}_{e^{2}}q+205625920q^{2}\nonumber\\
{\cal O}_{e}\cdot{\cal O}_{e^{4}}&=& {\cal O}_{e^{5}}
+300167{\cal O}_{e^{3}}q
+24699506832{\cal O}_{e}q^{2}\nonumber\\
{\cal O}_{e}\cdot{\cal O}_{e^{5}}&=&
{\cal O}_{e^{6}}
+200452{\cal O}_{e^{4}}q
+53751685624{\cal O}_{e^{2}}q^{2}\nonumber\\
&&+534155202302400q^{3}\nonumber\\
{\cal O}_{e}\cdot{\cal O}_{e^{6}}&=&{\cal O}_{e^{7}}
+56196{\cal O}_{e^{5}}q
+24699506832{\cal O}_{e^{3}}q^{2}\nonumber\\
&&+1920365635990032{\cal O}_{e}q^{3}\nonumber\\
{\cal O}_{e}\cdot{\cal O}_{e^{7}}&=&
5040{\cal O}_{e^{6}}q
+2056259520{\cal O}_{e^{4}}q^{2}\nonumber\\
&&+534155202302400{\cal O}_{e^{2}}q^{3}\nonumber\\
&&+5112982794486067200q^{4}\nonumber\\
\label{d13}
\end{eqnarray}
\begin{eqnarray}
H_{q,e}^{*}(M_{10}^{7})&&\nonumber\\
{\cal O}_{e}\cdot{\cal O}_{e}&=&{\cal O}_{e^{2}}\nonumber\\
{\cal O}_{e}\cdot{\cal O}_{e^{2}}&=&{\cal O}_{e^{3}}+5040q\nonumber\\
{\cal O}_{e}\cdot{\cal O}_{e^{3}}&=&{\cal O}_{e^{4}}
+56196{\cal O}_{e}q\nonumber\\
{\cal O}_{e}\cdot{\cal O}_{e^{4}}&=&{\cal O}_{e^{5}}
+200452{\cal O}_{e^{2}}q\nonumber\\
{\cal O}_{e}\cdot{\cal O}_{e^{5}}&=& {\cal O}_{e^{6}}
+300167{\cal O}_{e^{3}}q
+2091962880q^{2}\nonumber\\
{\cal O}_{e}\cdot{\cal O}_{e^{6}}&=&
{\cal O}_{e^{7}}
+200452{\cal O}_{e^{4}}q
+13570681320{\cal O}_{e}q^{2}\nonumber\\
{\cal O}_{e}\cdot{\cal O}_{e^{7}}&=&{\cal O}_{e^{8}}
+56196{\cal O}_{e^{5}}q
+13570681320{\cal O}_{e^{2}}q^{2}\nonumber\\
{\cal O}_{e}\cdot{\cal O}_{e^{8}}&=&
5040{\cal O}_{e^{6}}q
+2091962880{\cal O}_{e^{3}}q^{2}\nonumber\\
&&+13462263763200q^{3}\nonumber\\
\label{d14}
\end{eqnarray}
\newpage
\begin{eqnarray}
H_{q,e}^{*}(M_{11}^{8})&&\nonumber\\
{\cal O}_{e}\cdot{\cal O}_{e}&=&{\cal O}_{e^{2}}\nonumber\\
{\cal O}_{e}\cdot{\cal O}_{e^{2}}&=&{\cal O}_{e^{3}}+40320q\nonumber\\
{\cal O}_{e}\cdot{\cal O}_{e^{3}}&=&{\cal O}_{e^{4}}
+554112{\cal O}_{e}q\nonumber\\
{\cal O}_{e}\cdot{\cal O}_{e^{4}}&=&{\cal O}_{e^{5}}
+2552192{\cal O}_{e^{2}}q\nonumber\\
{\cal O}_{e}\cdot{\cal O}_{e^{5}}&=& {\cal O}_{e^{6}}
+5241984{\cal O}_{e^{3}}q
+345655618560q^{2}\nonumber\\
{\cal O}_{e}\cdot{\cal O}_{e^{6}}&=&
{\cal O}_{e^{7}}
+5241984{\cal O}_{e^{4}}q
+3857214283776{\cal O}_{e}q^{2}\nonumber\\
{\cal O}_{e}\cdot{\cal O}_{e^{7}}&=&{\cal O}_{e^{8}}
+2552192{\cal O}_{e^{5}}q
+8150222448640{\cal O}_{e^{2}}q^{2}\nonumber\\
{\cal O}_{e}\cdot{\cal O}_{e^{8}}&=&{\cal O}_{e^{9}}
+554112{\cal O}_{e^{6}}q
+3857214283776{\cal O}_{e^{3}}q^{2}\nonumber\\
&&+235354398279598080q^{3}\nonumber\\
{\cal O}_{e}\cdot{\cal O}_{e^{9}}&=&
40320{\cal O}_{e^{7}}q
+345655618560{\cal O}_{e^{4}}q^{2}\nonumber\\
&&+235354398279598080q^{3}{\cal O}_{e} \nonumber\\
\label{d15}
\end{eqnarray}
\newpage

\section*{\bf Appendix B  One Variable Polynomial Representation of 
$H^{*}_{q,e}(M_{N}^{k})$}
\begin{eqnarray}
H^{*}_{q,e}(M_{N}^{1})&&\nonumber\\
f_{rel}(X)&=& X^{N-1}-q\nonumber\\
{\cal O}_{e^{\alpha}}&=&X^{\alpha}\quad(0\leq\alpha\leq N-2)
\label{e1}
\end{eqnarray}
\begin{eqnarray}
H^{*}_{q,e}(M_{N}^{2})&&\nonumber\\
f_{rel}(X)&=& X^{N-1}-2^{2}Xq\nonumber\\
{\cal O}_{e^{\alpha}}&=&X^{\alpha}\quad(0\leq\alpha\leq N-3)\nonumber\\
{\cal O}_{e^{N-2}}&=&X^{N-2}-2qX
\label{e2}
\end{eqnarray}
\begin{eqnarray}
H^{*}_{q,e}(M_{N}^{3})&&\nonumber\\
f_{rel}(X)&=& X^{N-1}-3^{3}X^{2}q\nonumber\\
{\cal O}_{e^{\alpha}}&=&X^{\alpha}\quad(0\leq\alpha\leq N-4)\nonumber\\
{\cal O}_{e^{N-3}}&=&X^{N-3}-6q\nonumber\\
{\cal O}_{e^{N-2}}&=&X^{N-2}-21Xq
\label{e2a}
\end{eqnarray}
\begin{eqnarray}
H^{*}_{q,e}(M_{6}^{4})&&\nonumber\\
f_{rel}(X)&=& X^{5}-4^{4}X^{3}q\nonumber\\
{\cal O}_{e^{0}}&=&1\nonumber\\
{\cal O}_{e}&=&X\nonumber\\
{\cal O}_{e^{2}}&=&X^{2}-24q\nonumber\\
{\cal O}_{e^{3}}&=&X^{3}-128Xq\nonumber\\
{\cal O}_{e^{4}}&=&X^{4}-232X^{2}q-288q^{2}
\label{e3}
\end{eqnarray}
\begin{eqnarray}
H^{*}_{q,e}(M_{N}^{4})&(N\geq 7)&\nonumber\\
f_{rel}(X)&=& X^{N-1}-4^{4}X^{3}q\nonumber\\
{\cal O}_{e^{\alpha}}&=&X^{\alpha}\quad(0\leq\alpha\leq N-5)\nonumber\\
{\cal O}_{e^{N-4}}&=&X^{N-4}-24q\nonumber\\
{\cal O}_{e^{N-3}}&=&X^{N-3}-128Xq\nonumber\\
{\cal O}_{e^{N-2}}&=&X^{N-2}-232X^{2}q
\label{e4}
\end{eqnarray}
\newpage
\begin{eqnarray}
H^{*}_{q,e}(M_{7}^{5})&&\nonumber\\
f_{rel}(X)&=& X^{6}-5^{5}X^{4}q\nonumber\\
{\cal O}_{e^{0}}&=&1\nonumber\\
{\cal O}_{e}&=&X\nonumber\\
{\cal O}_{e^{2}}&=&X^{2}-120q\nonumber\\
{\cal O}_{e^{3}}&=&X^{3}-890Xq\nonumber\\
{\cal O}_{e^{4}}&=&X^{4}-2235X^{2}q-49800q^{2}\nonumber\\
{\cal O}_{e^{5}}&=&X^{5}-3005X^{3}q-57000Xq^{2}\nonumber\\
\label{e5}
\end{eqnarray}
\begin{eqnarray}
H^{*}_{q,e}(M_{8}^{5})&&\nonumber\\
f_{rel}(X)&=& X^{7}-5^{5}X^{4}q\nonumber\\
{\cal O}_{e^{0}}&=&1\nonumber\\
{\cal O}_{e}&=&X\nonumber\\
{\cal O}_{e^{2}}&=&X^{2}\nonumber\\
{\cal O}_{e^{3}}&=&X^{3}-120q\nonumber\\
{\cal O}_{e^{4}}&=&X^{4}-890Xq\nonumber\\
{\cal O}_{e^{5}}&=&X^{5}-2235X^{2}q\nonumber\\
{\cal O}_{e^{6}}&=&X^{6}-3005X^{3}q-7200q^{2}\nonumber\\
\label{e6}
\end{eqnarray}
\begin{eqnarray}
H^{*}_{q,e}(M_{N}^{5})&(N\leq 9)&\nonumber\\
f_{rel}(X)&=& X^{N-1}-5^{5}X^{4}q\nonumber\\
{\cal O}_{e^{\alpha}}&=&X^{\alpha}\quad 
(0\leq\alpha\leq N-6)\nonumber\\
{\cal O}_{e^{N-5}}&=&X^{N-5}-120q\nonumber\\
{\cal O}_{e^{N-4}}&=&X^{N-4}-890Xq\nonumber\\
{\cal O}_{e^{N-3}}&=&X^{N-3}-2235X^{2}q\nonumber\\
{\cal O}_{e^{N-2}}&=&X^{N-2}-3005X^{3}q\nonumber\\
\label{e7}
\end{eqnarray}
\newpage
\begin{eqnarray}
H^{*}_{q,e}(M_{8}^{6})&&\nonumber\\
f_{rel}(X)&=& X^{7}-6^{6}X^{5}q\nonumber\\
{\cal O}_{e^{0}}&=&1\nonumber\\
{\cal O}_{e}&=&X\nonumber\\
{\cal O}_{e^{2}}&=&X^{2}-720q\nonumber\\
{\cal O}_{e^{3}}&=&X^{3}-6984Xq\nonumber\\
{\cal O}_{e^{4}}&=&X^{4}-23328X^{2}q-7076160q^{2}\nonumber\\
{\cal O}_{e^{5}}&=&X^{5}-39672X^{3}q-24388128Xq^{2}\nonumber\\
{\cal O}_{e^{6}}&=&X^{6}-45936X^{4}q-9720000X^{2}q^{2}\nonumber\\
&&-5094835200q^{3}
\label{e8}
\end{eqnarray}
\begin{eqnarray}
H^{*}_{q,e}(M_{9}^{6})&&\nonumber\\
f_{rel}(X)&=& X^{8}-6^{6}X^{5}q\nonumber\\
{\cal O}_{e^{\alpha}}&=&X^{\alpha}\quad
(1\leq\alpha\leq 2)\nonumber\\
{\cal O}_{e^{3}}&=&X^{3}-720q\nonumber\\
{\cal O}_{e^{4}}&=&X^{4}-6984Xq\nonumber\\
{\cal O}_{e^{5}}&=&X^{5}-23328X^{2}q\nonumber\\
{\cal O}_{e^{6}}&=&X^{6}-39672X^{3}q-2384640q^{2}\nonumber\\
{\cal O}_{e^{6}}&=&X^{7}-45936X^{4}q-2643840Xq^{2}\nonumber\\
\label{e9}
\end{eqnarray}
\begin{eqnarray}
H^{*}_{q,e}(M_{10}^{6})&&\nonumber\\
f_{rel}(X)&=& X^{9}-6^{6}X^{5}q\nonumber\\
{\cal O}_{e^{\alpha}}&=&X^{\alpha}\quad
(1\leq\alpha\leq 3)\nonumber\\
{\cal O}_{e^{4}}&=&X^{4}-720q\nonumber\\
{\cal O}_{e^{5}}&=&X^{5}-6984Xq\nonumber\\
{\cal O}_{e^{6}}&=&X^{6}-23328X^{2}q\nonumber\\
{\cal O}_{e^{7}}&=&X^{7}-39672X^{3}q\nonumber\\
{\cal O}_{e^{6}}&=&X^{8}-45936X^{4}q-259200Xq^{2}\nonumber\\
\label{e10}
\end{eqnarray}
\newpage
\begin{eqnarray}
H^{*}_{q,e}(M_{9}^{7})&&\nonumber\\
f_{rel}(X)&=& X^{8}-7^{7}X^{6}q\nonumber\\
{\cal O}_{e^{0}}&=&1\nonumber\\
{\cal O}_{e}&=&X\nonumber\\
{\cal O}_{e^{2}}&=&X^{2}-5040q\nonumber\\
{\cal O}_{e^{3}}&=&X^{3}-61236Xq\nonumber\\
{\cal O}_{e^{4}}&=&X^{4}-261688X^{2}q-1045981440q^{2}\nonumber\\
{\cal O}_{e^{5}}&=&X^{5}-561855X^{3}q-7364461860Xq^{2}\nonumber\\
{\cal O}_{e^{6}}&=&X^{6}-762307X^{4}q
-8660264508X^{2}q^{2}\nonumber\\
&&-53577635146560q^{3}\nonumber\\
{\cal O}_{e^{7}}&=&X^{7}-818503X^{5}q
-1785767760X^{3}q^{2}\nonumber\\
&&-47590972087680q^{3}
\label{e11}
\end{eqnarray}
\begin{eqnarray}
H^{*}_{q,e}(M_{10}^{7})&&\nonumber\\
f_{rel}(X)&=& X^{9}-7^{7}X^{6}q\nonumber\\
{\cal O}_{e^{0}}&=&1\nonumber\\
{\cal O}_{e}&=&X\nonumber\\
{\cal O}_{e^{2}}&=&X^{2}\nonumber\\
{\cal O}_{e^{3}}&=&X^{3}-5040q\nonumber\\
{\cal O}_{e^{4}}&=&X^{4}-61236Xq\nonumber\\
{\cal O}_{e^{5}}&=&X^{5}-261688X^{2}q\nonumber\\
{\cal O}_{e^{6}}&=&X^{6}-561855X^{3}q-579121200q^{2}\nonumber\\
{\cal O}_{e^{7}}&=&X^{7}-762307X^{4}q
-1874923848Xq^{2}\nonumber\\
{\cal O}_{e^{8}}&=&X^{8}-818503X^{5}q
-739786320X^{2}q^{2}
\label{e12}
\end{eqnarray}
\newpage
\begin{eqnarray}
H^{*}_{q,e}(M_{11}^{8})&&\nonumber\\
f_{rel}(X)&=& X^{10}-8^{8}X^{7}q\nonumber\\
{\cal O}_{e^{0}}&=&1\nonumber\\
{\cal O}_{e}&=&X\nonumber\\
{\cal O}_{e^{2}}&=&X^{2}\nonumber\\
{\cal O}_{e^{3}}&=&X^{3}-40320q\nonumber\\
{\cal O}_{e^{4}}&=&X^{4}-594432Xq\nonumber\\
{\cal O}_{e^{5}}&=&X^{5}-3146624X^{2}q\nonumber\\
{\cal O}_{e^{6}}&=&X^{6}-8388608X^{3}q-134298823680q^{2}\nonumber\\
{\cal O}_{e^{7}}&=&X^{7}-13630592X^{4}q
-875510074368Xq^{2}\nonumber\\
{\cal O}_{e^{8}}&=&X^{8}-16182784X^{5}q
-994943923200X^{2}q^{2}\nonumber\\
{\cal O}_{e^{9}}&=&X^{9}-16736896X^{6}q
-203929850880X^{3}q^{2}\nonumber\\
&&-5414928570777600q^{3}
\label{e11a}
\end{eqnarray}

\end{document}